\documentclass[%
groupedaddress,
preprint,
showpacs,
 amsmath,amssymb,
 aps,
prb,
]{revtex4-1}

\newcommand\beq{\begin{equation}}
\newcommand\eeq{\end{equation}}

\usepackage{graphicx}
\usepackage{dcolumn}
\usepackage{bm}

\begin{document}


\title{Optical Nonlocality in Multilayered Hyperbolic Metamaterials Based on Thue-Morse Superlattices}

\author{Silvio Savoia}
\author{Giuseppe Castaldi}
\author{Vincenzo Galdi}
\email{vgaldi@unisannio.it}
\affiliation{Waves Group, Department of Engineering, University of Sannio, I-82100 Benevento, Italy
}%

\date{\today}


\begin{abstract}
We show that hyperbolic electromagnetic metamaterials implemented as multilayers based on two material constituents arranged according to Thue-Morse (ThM) aperiodic sequence may exhibit strong nonlocal effects, manifested as the appearance of additional extraordinary waves which are not predicted by standard  effective-medium-theory (local) models. From the mathematical viewpoint, these effects can be associated with stationary points of the transfer-matrix trace, and can be effectively parameterized via the trace-map formalism. We show that their onset is accompanied by a strong wavevector dependence in the effective constitutive parameters. In spite of the inherent periodicity enforced by the unavoidable (Bloch-type) supercell terminations, we show that such strong nonlocality is retained at any arbitrarily high-order iterations, i.e., approaching the actual aperiodic regime. Moreover, for certain parameter configurations, at a given wavelength and for two given material layers, these effects may be significantly less prominent when the same layers are arranged in a standard periodic fashion. Our findings indicate that the (aperiodic) positional order of the layers constitutes an effective and technologically inexpensive additional degree of freedom in the engineering of optical nonlocality.
\end{abstract}

\maketitle

\section{Introduction}

Electromagnetic (EM) metamaterials are artificial materials composed of subwavelength dielectric and/or metallic inclusions in a host medium, which have attracted considerable scientific and application-oriented attention due to possibility
to engineer anomalous properties (e.g., negative refraction) that are not observable in natural materials.\cite{Capolino2009} Of particular interest are the so-called ``hyperbolic'' metamaterials,\cite{Smith2004,Noginov2009} characterized by nonmagnetic, uniaxally-anisotropic constitutive relationships with both positive and negative components of the permittivity tensor. This yields a {\em hyperbolic} (as opposed to {\em spherical}, in conventional isotropic media) dispersion relationship, which allows for propagation of (otherwise evanescent) waves with large wavevectors, resulting in a high (theoretically unbounded) photonic density of states. The reader is referred to Refs. \onlinecite{Jacob2006,Govyadinov2006,Smolyaninov2010,Jacob2010,Yao2011,Krishnamoorthy2012,Jacob2012,Cortes2012,Biehs2012} for a sparse sampling of applications, ranging from nanoimaging to quantum nanophotonics and thermal emission.

In what follows, we focus on multilayered hyperbolic metamaterials,\cite{Jacob2010} implemented via stacking of alternating subwavelength layers with negative and positive permittivities (e.g., metallic and dielectric, at optical wavelengths). For this class, the effective medium theory (EMT) provides a particularly simple model in terms of a homogeneous, uniaxially-anisotropic permittivity tensor with components given by the Maxwell-Garnett mixing formulas.\cite{Sihvola1999} However, a series of recent papers \cite{Elser2007,Orlov2011,Chebykin2012,Kidwai2012,Orlov2013} have pointed out the limitations of this model in predicting {\em nonlocal} effects that can take place (even in the presence of deep subwavelength layers) due to the coupling of surface plasmon polaritons (SPPs) propagating along the interfaces separating layers with oppositely-signed permittivities. This may result, for instance, in the misprediction of additional extraordinary waves \cite{Orlov2011,Orlov2013} as well as of the broadband Purcell effect.\cite{Kidwai2012}


We point out that typical multilayered hyperbolic metamaterials are based on {\em periodic} arrangements of the layers.\cite{Jacob2010} In fact, the EMT model describing the {\em local} response is independent of the positional order of the layers, and depends only on the permittivities of the two constituents and their filling fractions.\cite{Sihvola1999} However, one would intuitively expect the positional order of the layers to sensibly affect the {\em nonlocal} response.
It seems therefore suggestive to investigate possible nonlocal effects in structures characterized by {\em aperiodic order} inspired by the ``quasicrystal'' concept in solid-state physics.\cite{Shechtman1984,Levine1984}

Within this framework, we study here the nonlocal response of hyperbolic metamaterials constituted by multilayer superlattices based on the  Thue-Morse (ThM) geometry.\cite{Queffelec2010} For these structures, we identify certain nonlocal effects, in terms of additional extraordinary waves, that are strongly dependent on the specific positional order of the material layers. Accordingly, the rest of the paper is laid out as follows. In Sec. \ref{Sec:Formulation}, we introduce the problem geometry and formulation, as well as the main analytical tools utilized (with details relegated in Appendices \ref{Sec:APPA} and \ref{Sec:APPB}). In Sec. \ref{Sec:Results}, we derive, illustrate and validate the main results, via analytical studies, retrieval of effective wavevector-dependent constitutive parameters, and full-wave numerical simulations. In Sec. \ref{Sec:Remarks} and  Sec. \ref{Sec:Conclusions}, we provide some additional remarks and conclusions, respectively.

\section{Problem Formulation and Analytical Modeling}
\label{Sec:Formulation}

\subsection{Problem Geometry and Generalities}

\begin{figure}
\begin{center}
\includegraphics [width=10cm]{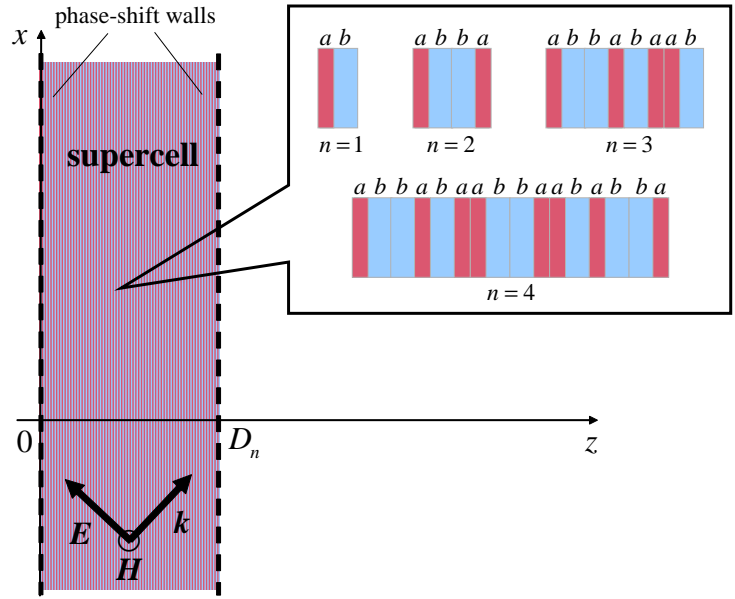}
\end{center}
\caption{(Color online) Problem schematic, illustrating the 2-D propagation of TM-polarized EM fields (with $y$-directed magnetic field)
in the ThM-based hyperbolic metamaterial of interest (details in the text).}
\label{Figure1}
\end{figure}

Referring to the two-dimensional (2-D) $y-$independent geometry in Fig. \ref{Figure1}, we consider a multilayer superlattice obtained via the infinite repetition along the $z-$axis of a supercell composed of layers of two nonmagnetic, material constituents labeled with the letters ``$a$'' and ``$b$'' (with relative permittivities $\varepsilon_a$ and $\varepsilon_b$, and thicknesses $d_a$ and $d_b$, respectively), arranged according to the ThM sequence. Assuming as an initiator the sequence ``$ab$,'' this amounts to iterating the following inflation rules \cite{Queffelec2010}
\beq 
a\rightarrow ab,~~b\rightarrow ba,
\label{eq:SubRule}
\eeq
as shown schematically in the inset of Fig. \ref{Figure1} for the first iteration-orders $n$. In what follows, we study the time-harmonic  [$\exp(-i\omega t)$] propagation of transversely-magnetic (TM) polarized EM fields, neglecting for now material losses, as previous studies \cite{Orlov2013} have shown that they only mildly affect optical nonlocality.

It is readily recognized that the first two iterations ($n=1,2$) correspond to standard periodic multilayers (with period $d=d_a+d_b$ and $2d$, respectively); these will be accordingly referred to as ``standard periodic'' cases. In fact, we emphasize that the geometry in Fig. \ref{Figure1} is {\em inherently periodic} for any finite iteration-order $n$, and approaches the actual aperiodic regime in the limit $n\rightarrow\infty$ (see Sec. \ref{Sec:Aperiodic} below). Moreover, we observe that, given the structure of the inflation rule in (\ref{eq:SubRule}), any iteration-order of our ThM multilayer contains the same proportions of ``$a$''-type and ``$b$''-type constituents  as the standard periodic case, and differs solely in the positional order of the layers. Accordingly, the Maxwell-Garnett mixing formulas for the parallel ($\parallel$, i.e., $x,y$) and orthogonal ($\perp$, i.e., $z$) permittivity components \cite{Sihvola1999}
\beq
\varepsilon_{\parallel}=\frac{\varepsilon_a d_a+\varepsilon_b d_b}{d},~~
\varepsilon_{\perp}=\left(\frac{\varepsilon_a^{-1} d_a+\varepsilon_b^{-1} d_b}{d}\right)^{-1},
\label{eq:MG}
\eeq
yield the same EMT model for any iteration-order, which results in the dispersion relationship
\beq
\frac{k_x^2}{\varepsilon_{\perp}}+\frac{k_z^2}{\varepsilon_{\parallel}}=k^2,
\label{eq:disp}
\eeq
where $k_x$ and $k_z$ indicate the $x-$ and $z-$ components, respectively,  of the wavevector ${\bf k}$ (cf. Fig. \ref{Figure1}), and $k=\omega/c=2\pi/\lambda$ indicates the vacuum wavenumber (with $c$ and $\lambda$ denoting the corresponding wavespeed and wavelength). By suitably choosing the parameters in the mixing rules (\ref{eq:MG}) so that $\varepsilon_{\parallel}\varepsilon_{\perp}<0$, the dispersion relationship in (\ref{eq:disp}), interpreted in terms of {\em equi-frequency contour} (EFC), assumes the anticipated {\em hyperbolic} character.
Since the local EMT model in (\ref{eq:MG}) and (\ref{eq:disp}) is {\em identical} for any iteration-order, any possible difference in the (nonlocal) responses should solely be attributed to the different positional order of the material layers.

\subsection{Exact Dispersion Relationship}

Multilayers based on the ThM geometry have been widely studied in the past, in the form of dielectric/semiconductor photonic quasicrystals (see, e.g., Refs. \onlinecite{Liu1997,Qiu2003,DalNegro2004,Grigoriev2010,Hsueh2011,Jiang2005} for a sparse sampling), and with main focus on the resonant-transmission, localization, omnidirectional-reflection, and bandgap properties. To the best of our knowledge, no previous attempt was made to study ThM-based hyperbolic metamaterials. Following a rather standard approach (see Appendix \ref{Sec:APPA} for more details), the {\em exact} dispersion relationship pertaining to a $n$th-order ThM supercell terminated by Bloch-type phase-shift walls (cf. Fig. \ref{Figure1}) can be compactly written as
\beq
\cos\left(
k_z D_n
\right)=\frac{\chi_n}{2},
\label{eq:Bloch-Disp}
\eeq
where $D_n=2^{n-1}d$ represents the total supercell thickness at the iteration-order $n$, and $\chi_n$ denotes the {\em trace} (i.e., the sum of the diagonal elements) of the transfer-matrix that relates the tangential components of the EM fields at the supercell interfaces $z=0$ and $z=D_n$. For the first two iterations $n=1,2$, the trace can be straightforwardly calculated as 
\beq
\chi_n\!=\!2\cos\!\left(n
\delta_a\right)
\!\cos\!\left(n
\delta_b\right)
\!-\!\left(
\frac{\gamma_a}{\gamma_b}+\frac{\gamma_b}{\gamma_a}
\right)
\!\sin\!\left(n
\delta_a\right)\!\sin\!\left(n
\delta_b\right),
\label{eq:chi12}
\eeq
thereby recovering the familiar Bloch-type dispersion relationship of standard periodic multilayers (as in Refs. \onlinecite{Elser2007,Orlov2011,Chebykin2012,Kidwai2012,Orlov2013}), where
\beq
\delta_{a,b}=k_{za,b}d_{a,b},~~\gamma_{a,b}=\frac{\varepsilon_{a,b}k}{k_{za,b}},
\eeq
with $k_{za,b}=\sqrt{k^2\varepsilon_{a,b}-k_x^2}$, $\mbox{Im}(k_{za,b})\ge 0$.
For higher-order iterations, a particularly simple recursive calculation procedure can be adopted, based on the {\em trace-map} \cite{Kolar1990,Wang2000} (see also Appendix \ref{Sec:APPB} for details)
\beq
\chi_{n+2}=\chi_n^2\left(
\chi_{n+1}-2
\right)+2,~~n\ge 1.
\label{eq:tracemap}
\eeq

\subsection{Conditions for Additional Extraordinary Waves}

The exact dispersion relationship in (\ref{eq:Bloch-Disp}) reduces to the local EMT model in (\ref{eq:disp}) in the limit $d\rightarrow 0$, but it may significantly depart from that for finite (and yet subwavelength) layer thicknesses. In particular, we are interested in exploring possible nonlocal effects manifested as the appearance of additional extraordinary waves that are not predicted by the local EMT model in (\ref{eq:disp}), and have already been observed in hyperbolic metamaterials based on standard periodic multilayers. \cite{Orlov2011,Orlov2013}
From the mathematical viewpoint, this phenomenon is related to {\em multiple} (apart from sign) $k_x$ solutions of (\ref{eq:Bloch-Disp}) for a given value of $k_z$ and $\omega$, which may occur if the trace $\chi_n$ is a {\em nonmonotonic} function of $k_x^2$. We are therefore led to study the {\em stationary points} of the trace $\chi_n$ with respect to the argument $k_x^2$,
\beq
{\dot \chi}_n\left(k_x^2\right)=0.
\eeq

\section{Representative Results}
\label{Sec:Results}
\subsection{Trace-Map and EFC Studies}

In order to better emphasize the role of the positional order in the onset of these nonlocal phenomena, we deliberately focus on parameter configurations for which the hyperbolic metamaterial arising from the first iteration $n=1$ (i.e., a standard periodic multilayer) is well-described by the local EMT model in (\ref{eq:disp}). This translates in the trace $\chi_1$ in (\ref{eq:chi12}) being well approximated by its second-order Taylor expansion in $d$,
\beq
\chi_1\left(k_x^2\right)\approx 2-\varepsilon_{\parallel}k^2d^2+\frac{\varepsilon_{\parallel}d^2}{\varepsilon_{\perp}} k_x^2.
\label{eq:chi1app}
\eeq

\begin{figure}
\begin{center}
\includegraphics [width=12cm]{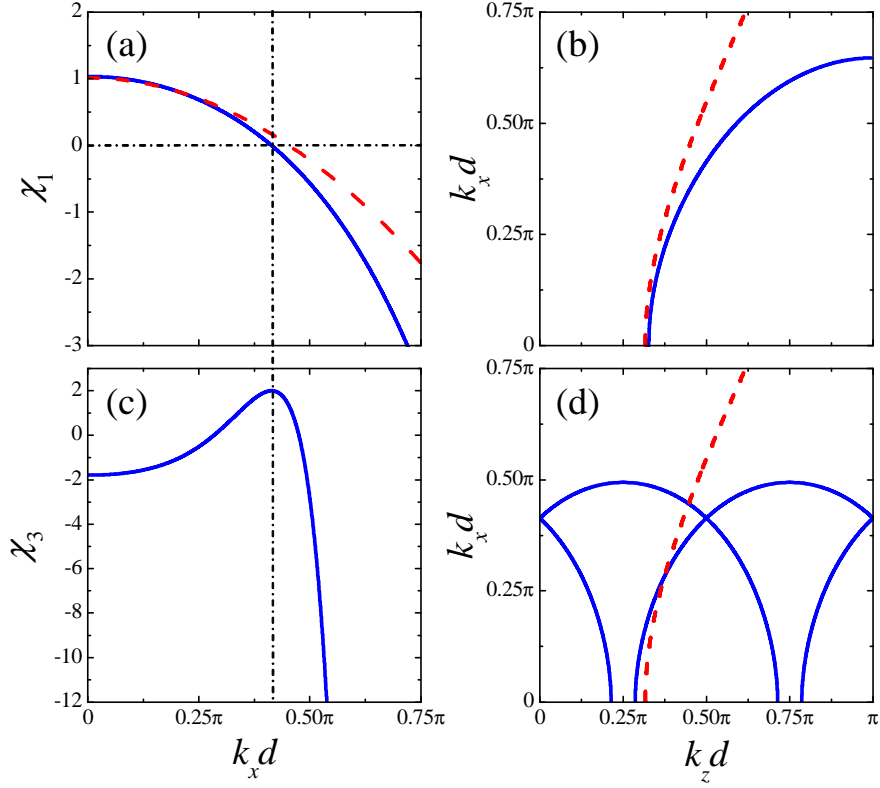}
\end{center}
\caption{(Color online) (a) Transfer-matrix trace (blue-solid curve) as a function of $k_x$ and (b) corresponding EFC (within the first Brilluoin zone), for a hyperbolic metamaterial with $\varepsilon_a=6.83$, $\varepsilon_b=-1.83$, and $d_a=d_b=d/2=0.05\lambda$ (i.e., $\varepsilon_{\parallel}=2.5$, $\varepsilon_{\perp}=-5$) at the $n=1$ iteration-order (i.e., standard periodic multilayer of period $d$). Also shown (red-dashed curves) are the predictions from the local EMT model. Due to symmetry, only positive values of $k_x$ and $k_z$ are shown. (c), (d) Same as above, but for the $n=3$ iteration-order (four Brillouin zones shown for direct comparison). The dash-dotted lines highlight the correspondence between the zero of $\chi_1$ and the maximum of $\chi_3$.}
\label{Figure2}
\end{figure}

Figure \ref{Figure2}(a) compares, for one such parameter configuration (given in the caption, and corresponding to $\varepsilon_{\parallel}=2.5$ and $\varepsilon_{\perp}=-5$), the exact trace $\chi_1$ [cf. (\ref{eq:chi12})] and its quadratic approximation (\ref{eq:chi1app}), showing a reasonable agreement. The corresponding exact [cf. (\ref{eq:Bloch-Disp})] EFCs and the local-EMT prediction [cf. (\ref{eq:disp})] are compared in Fig. \ref{Figure2}(b)
within the first Brillouin zone $0\le k_z\le \pi/d$. Again, we observe a good agreement (especially for smaller values of $k_z$) and, most important, a {\em single} branch. This yields a {\em single} mode that is propagating for $k_z\gtrsim\sqrt{\varepsilon_{\parallel}}k\approx0.32\pi/d$, and evanescent otherwise.
We now move on to looking at higher-order iterations of the ThM geometry. From the trace-map in (\ref{eq:tracemap}), we straightforwardly obtain
\beq
{\dot \chi}_{n+2}=\chi_n\left[
\chi_n{\dot \chi}_{n+1}+2{\dot \chi}_n\left(\chi_{n+1}-2\right)
\right],~~n\ge 1,
\label{eq:chi1pmap}
\eeq
with the overdot denoting differentiation with respect to the argument $k_x^2$. This implies that the vanishing of the trace at a given iteration-order, i.e., $\chi_n=0$ is a sufficient condition for a stationary point ${\dot \chi}_{n+2}=0$, and hence the presence of additional extraordinary waves, at a higher-order iteration. Therefore, if the trace $\chi_1$ admits a zero, then $\chi_3$ should exhibit at least one stationary point. For the assumed parameter configuration, for which the local EMT model, and hence the quadratic approximation in (\ref{eq:chi1app}), holds reasonably well at the first iteration-order, the position $k_{x0}$ of such zero (and corresponding stationary point) admits a simple analytical estimate as
\beq
k_{x0}\approx
\frac{1}{d}\sqrt{\frac{\varepsilon_{\perp}\left(\varepsilon_{\parallel}k^2d^2-2\right)}{\varepsilon_{\parallel}}},
\label{eq:kx0}
\eeq
which yields a real solution provided that $\varepsilon_{\parallel}k^2d^2\le 2$. In our case, this last condition is verified and, as can be observed from Fig. \ref{Figure2}(a), the estimate in (\ref{eq:kx0}) is moderately accurate, yielding a $9\%$ error with respect to the actual zero position $k_{x0}=0.4138 d/\pi$ calculated numerically. For the same parameter configuration, Fig. \ref{Figure2}(c) shows the trace $\chi_3$ at the $n=3$ iteration-order, from which a maximum at $k_{x0}$ can be observed. As a consequence, besides a branch that is still in good agreement with the local EMT prediction, the corresponding EFCs shown in Fig. \ref{Figure2}(d) [within a spectral region covering four Brilluoin zones, in order to facilitate direct comparison with Fig. \ref{Figure2}(b)] exhibit additional branches, resulting in two additional modes (extraordinary waves) that propagate for arbitrarily small values of $k_z$, and degenerate into a single mode at $k_z^{(m)}=m\pi/(2d)$, $m=0,1,2,..$.

\begin{figure}
\begin{center}
\includegraphics [width=12cm]{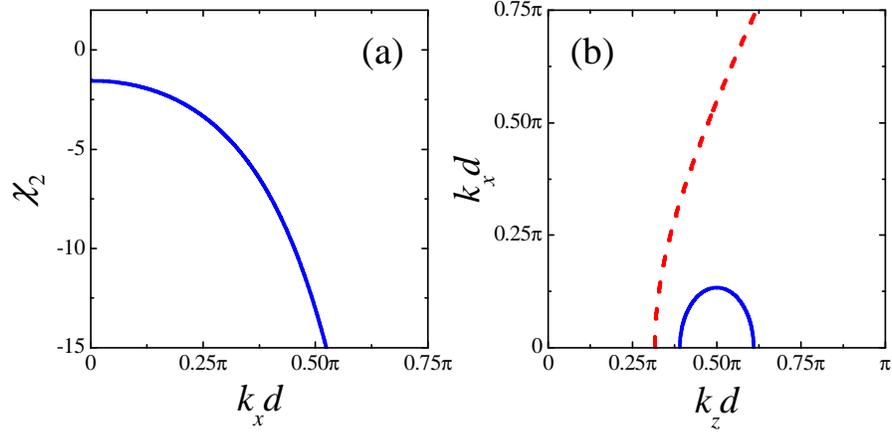}
\end{center}
\caption{(Color online) As is Figs. \ref{Figure2}(a) and \ref{Figure2}(b), but for the $n=2$ iteration-order (standard periodic multilayer of period $2d$; two Brillouin zones shown in the EFCs).}
\label{Figure3}
\end{figure}

\begin{figure}
\begin{center}
\includegraphics [width=12cm]{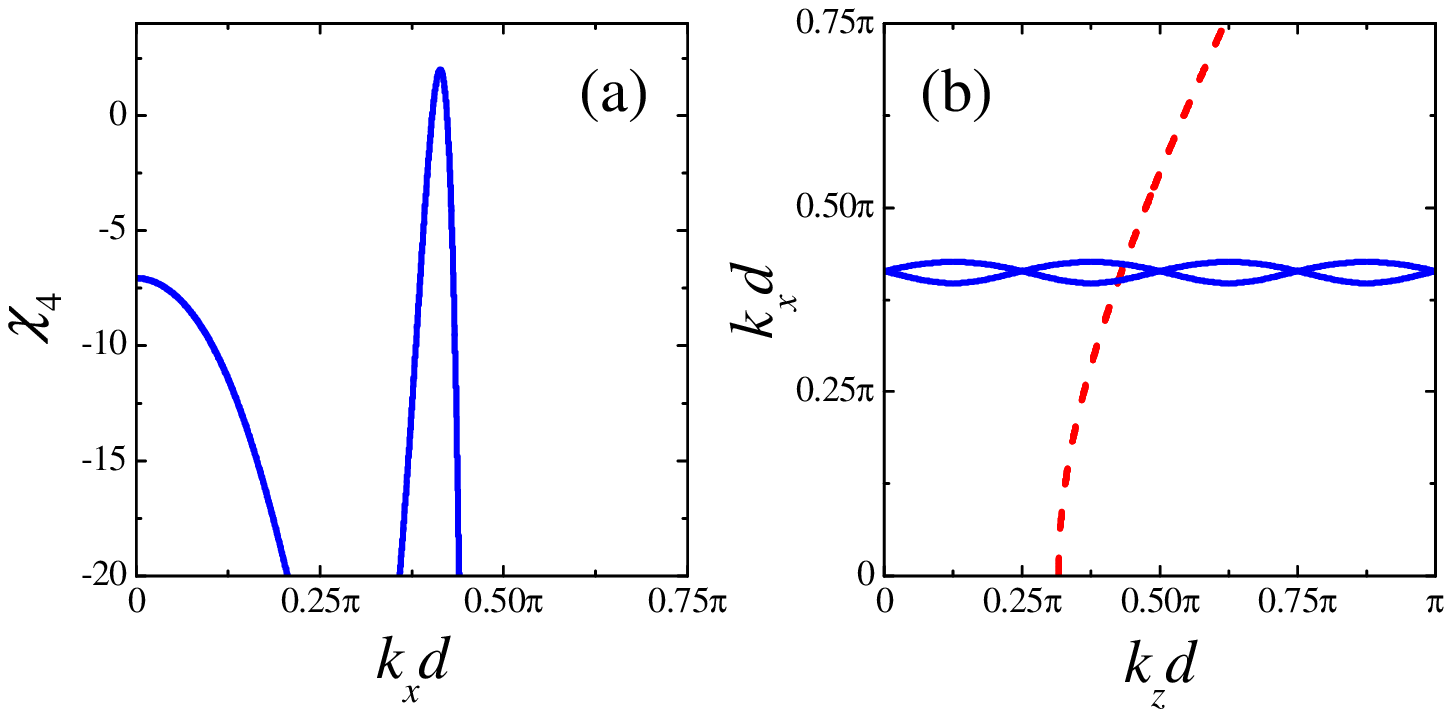}
\end{center}
\caption{(Color online) As is Figs. \ref{Figure2}(a) and \ref{Figure2}(b), but for the $n=4$ iteration-order (eight Brillouin zones shown in the EFCs).}
\label{Figure4}
\end{figure}

\begin{figure}
\begin{center}
\includegraphics [width=12cm]{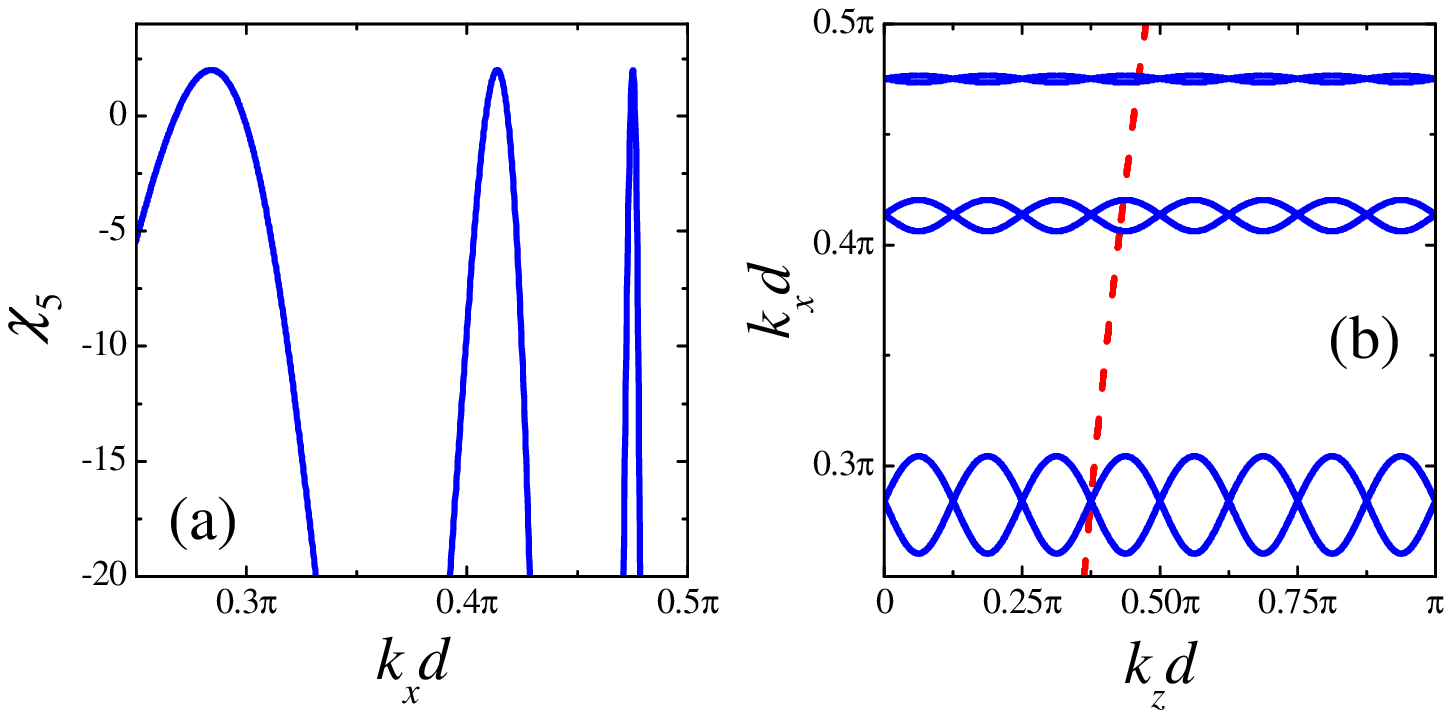}
\end{center}
\caption{(Color online) As is Figs. \ref{Figure2}(a) and \ref{Figure2}(b), but for the $n=5$ iteration-order (sixteen Brillouin zones shown in the EFCs).}
\label{Figure5}
\end{figure}

Figures \ref{Figure3}-\ref{Figure5} illustrate the results pertaining to the intermediate ($n=2$) and higher ($n=4$ and $n=5$) iteration-orders, respectively. More specifically, the $n=2$  iteration-order illustrated in Fig. \ref{Figure3} still corresponds to a standard periodic multilayer (of period $2d$). For the chosen parameter configuration, the local EMT prediction is less accurate than what observed for the   case [cf. Figs. \ref{Figure2}(a) and \ref{Figure2}(b)], but still correctly predicts a single mode that is evanescent below a cutoff value of  $k_z$. 
Likewise, looking at the results for the $n=4$  iteration-order in Fig. \ref{Figure4}, we note that the trace $\chi_4$ exhibits a maximum ($\chi_4=2$) at $k_{x0}$. We therefore obtain two (always propagating) modes which degenerate at $k_z^{(m)}=m\pi/(4d)$, $m=0,1,2,...$. This is qualitatively similar to what observed for the $n=3$ case [cf. Figs. \ref{Figure2}(c) and \ref{Figure2}(d)], although now both modes turn out to depart substantially from the local EMT prediction. 
Rather different, and quite interesting, are the results pertaining to the $n=5$ iteration-order. As can be observed from Fig. \ref{Figure5}(a), the trace $\chi_5$  now exhibits {\em three} maxima ($\chi_5=2$). In the corresponding EFCs [Fig. \ref{Figure5}(b)], this translates into {\em six} (always propagating) modes which degenerate at $k_z^{(m)}=m\pi/(8d)$, $m=0,1,2,...$

\subsection{Nonlocal Effective Constitutive Parameters}
\label{Sec:NLP}
An effective nonlocal model capable of capturing the above effects can be derived in terms of a homogeneous uniaxial medium with {\em wavevector-dependent} relative-permittivity components, ${\hat \varepsilon}_{\parallel}\left({\bf k}\right)$ and ${\hat \varepsilon}_{\perp}\left({\bf k}\right)$, whose dispersion law 
\beq
\frac{k_x^2}{{\hat \varepsilon}_{\perp}\left({\bf k}\right)}+\frac{k_z^2}{{\hat \varepsilon}_{\parallel}\left({\bf k}\right)}=k^2
\label{eq:dispNL}
\eeq
suitably approximates the exact dispersion law in (\ref{eq:Bloch-Disp}). 
In Ref. \onlinecite{Elser2007}, for a similar (standard periodic multilayer) configuration, such a model was successfully derived in terms of second-order rational functions (of $k_x$ and $k_z$), so that the exact and approximate dispersion laws would match up to the fourth order in $d$. In our case here, in view of the generally larger dynamical ranges observed, we found it necessary to derive a higher-order model,
\begin{subequations}
\begin{eqnarray}
{\hat \varepsilon}_{\parallel}\left(k_x\right)&=&\frac{1}{\alpha_0+\alpha_2 k_x^2+\alpha_4 k_x^4+\alpha_6 k_x^6+\alpha_8 k_x^8},
\label{eq:epar}\\
{\hat \varepsilon}_{\perp}\left(k_z\right)&=&\frac{1}{\beta_0+\beta_2 k_z^2+\beta_4 k_z^4+\beta_6 k_z^6+\beta_8 k_z^8},
\label{eq:eperp}
\end{eqnarray}
\label{eq:effNL}
\end{subequations}
by matching the exact dispersion law in (\ref{eq:Bloch-Disp}) up to the tenth-order in $d$. The coefficients $\alpha_l$ and $\beta_l$, $l=0,2,4,6,8,$ generally depend on the frequency and on the geometrical and constitutive parameters of the multilayer, and are given in Tables \ref{Table1} and \ref{Table2}, respectively, for the parameter configuration and iteration-orders ($n=1$ and $n=3$) as in Fig. \ref{Figure2}. 

\begin{figure}
\begin{center}
\includegraphics [width=12cm]{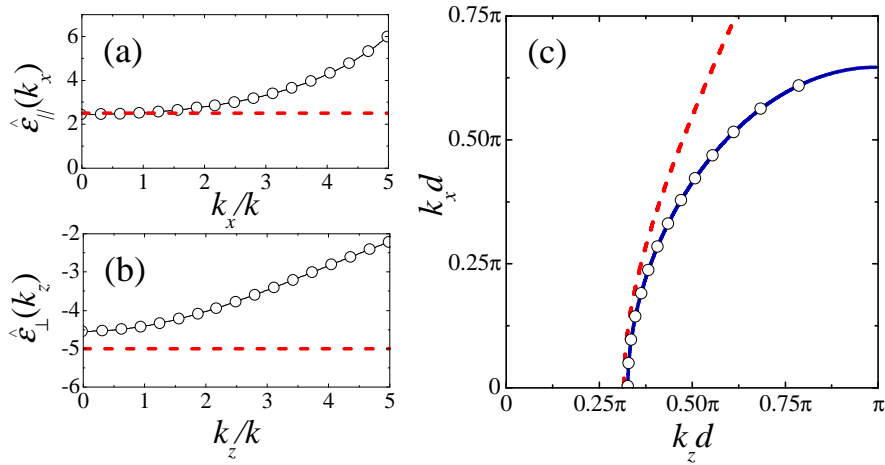}
\end{center}
\caption{(Color online) Geometry and parameters as in Fig. \ref{Figure2}. (a), (b) Effective nonlocal relative-permittivities ${\hat \varepsilon}_{\parallel}(k_x)$ and ${\hat \varepsilon}_{\perp}(k_z)$ [cf. (\ref{eq:effNL}) and Tables \ref{Table1} and \ref{Table2}], respectively, as a function of the normalized wavenumbers, for the $n=1$ iteration-order (circles). (c) Corresponding EFC [cf. (\ref{eq:dispNL})]. Also shown, as references, are the local EMT predictions (red-dashed curves) and the exact EFC (blue-solid curve).}
\label{Figure6}
\end{figure}

As can be expected, for the $n=1$ iteration-order (standard periodic multilayer), the coefficients pertaining to higher-order terms in $k_x$ and $k_z$ are quite small. The resulting mild wavevector-dependence in the effective parameters is also visible in Figs. \ref{Figure6}(a) and \ref{Figure6}(b), while Fig. \ref{Figure6}(c) shows the excellent agreement between the exact EFC and the prediction from the nonlocal effective model. 

\begin{figure}
\begin{center}
\includegraphics [width=12cm]{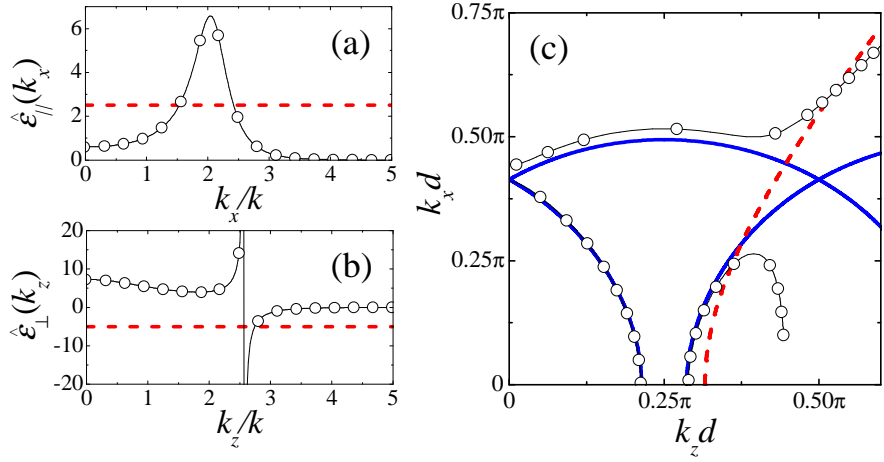}
\end{center}
\caption{(Color online) As in Fig. \ref{Figure6}, but for the $n=3$ iteration-order.}
\label{Figure7}
\end{figure}

Conversely, for the $n=3$ iteration-order, these higher-order coefficients are {\em non-negligible}, thereby confirming the quite strong  wavevector dependence in the effective constitutive parameters, as evidently visible in Fig. \ref{Figure7}(a) and \ref{Figure7}(b). Overall, as shown in Fig. \ref{Figure7}(c), the nonlocal effective model is capable of accurately capturing (over the first Brillouin zone) the peculiar multi-branch behavior of the EFCs.

It should be stressed that the above approach is not the only one available (see, e.g., Ref. \onlinecite{Chebykin2012} for an alternative), and that it is inherently limited to the modal analysis in a {\em bulk} medium scenario. Applications to boundary-value problems generally require more refined models as well as the derivation of additional boundary conditions.\cite{Maslovski2010}

\subsection{Propagation Through a Slab}

We now focus on an independent validation of our findings above. To this aim, for computational affordability, we study the TM plane-wave propagation through a slab of our ThM-based hyperbolic metamaterial (infinitely long in the $z-$direction and of finite thickness along $x$, with parameters as in Fig. \ref{Figure2}) immersed in vacuum, at various iteration-orders.
Our numerical simulations below rely on a Rigorous Coupled Wave Analysis (RCWA) algorithm,\cite{Moharam1995} based on the Fourier-series expansion of the piecewise-constant permittivity distribution of the supercell. In our study, the truncation of this expansion was chosen according to a convergence criterion based on the root-mean-square (RMS) variations of the magnetic-field distribution within the supercell. Basically, the number of modes in the expansion was increased until RMS variations $<2\%$ were observed. For the larger supercells considered in our study ($n=5$ iteration-order, i.e., 32 layers), convergence was achieved by using 391 modes.

\begin{figure}
\begin{center}
\includegraphics [width=12cm]{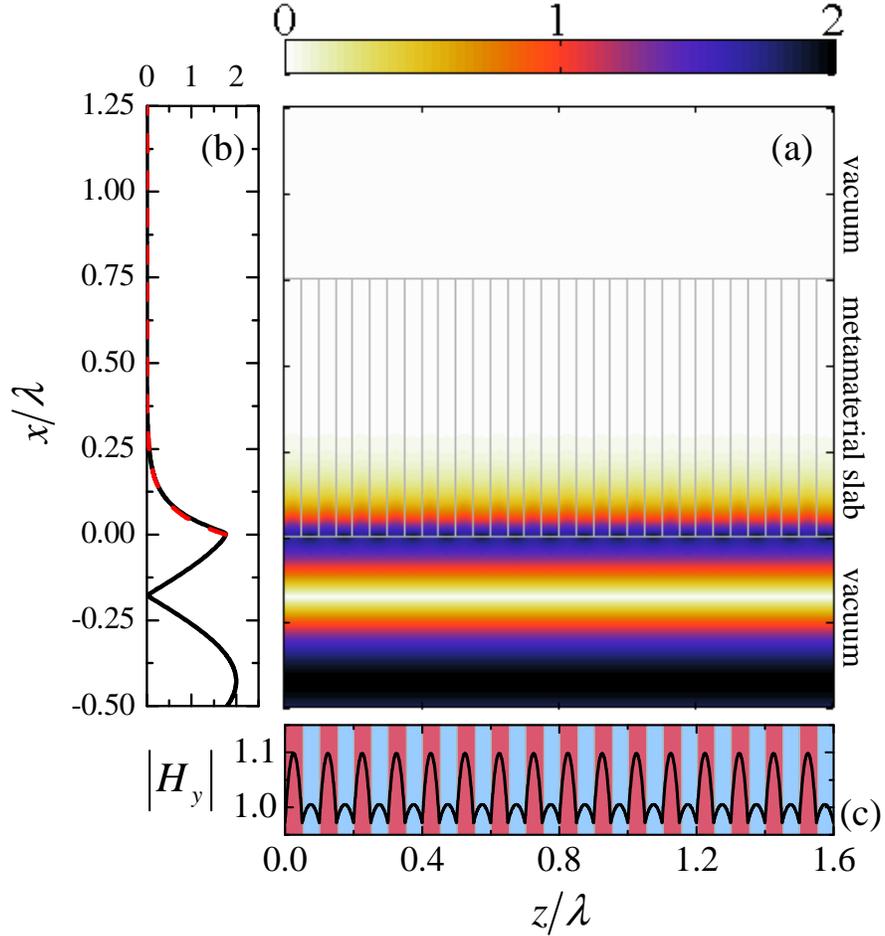}
\end{center}
\caption{(Color online) (a) Numerically-computed field-magnitude ($H_y$) map in false-color scale, for a slab of hyperbolic metamaterial (parameters as in Fig. \ref{Figure2}) at the $n=1$ iteration-order (standard periodic multilayer of period $d$; 16 unit cells shown) of thickness $L=0.75\lambda$, embedded in vacuum, and excited by a unit-amplitude, normally-incident ($k_z=0$) plane-wave. Thin grey lines delimit the slab and layer interfaces. (b) Longitudinal cut (black-solid curve) at $z=0.05\lambda$, and exponential fit (red-dashed curve) of the evanescent decay inside the slab as predicted by the local EMT model [cf. (\ref{eq:alphax})]. (c) Transverse cut at $x=0.05\lambda$, with the material layers visualized with different colors/shades.}
\label{Figure8}
\end{figure}

\begin{figure}
\begin{center}
\includegraphics [width=12cm]{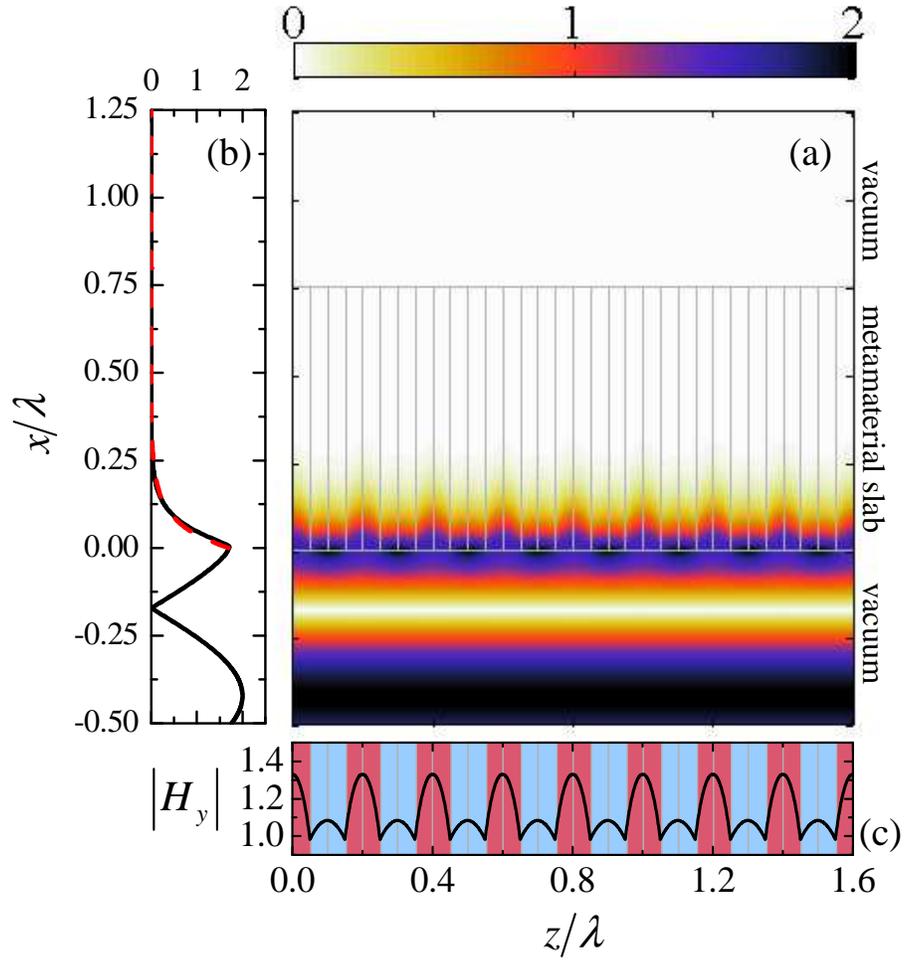}
\end{center}
\caption{(Color online) As in Fig. \ref{Figure3}, but for the $n=2$ iteration-order (standard periodic multilayer of period $2d$; eight unit-cells shown).}
\label{Figure9}
\end{figure}

Figure \ref{Figure8}(a) shows a field-magnitude map pertaining to the first iteration-order $n=1$ (i.e., standard periodic multilayer) for normal incidence ($k_z=0$) and thickness $L=0.75\lambda$. As can be observed also from the longitudinal ($x$) cut in Fig. \ref{Figure8}(b), the field is totally reflected, with only an evanescent decay inside the slab, which is accurately fitted by an exponential tail (red-dashed curve) with attenuation coefficient
\beq
\alpha_x\approx \sqrt{\left|\epsilon_{\perp}\right|} k, 
\label{eq:alphax}
\eeq
as predicted by the local EMT model in (\ref{eq:disp}). The $z$-cut in Fig. \ref{Figure8}(c) shows that the transverse field distribution is rather uniform ($\sim 10\%$ variations) and weakly peaked at the centers of the layers. Overall, as expected, the local EMT model provides a satisfactory prediction. Similar considerations hold for the $n=2$ iteration-order (standard periodic multilayer of period $2d$) shown in Fig. \ref{Figure9}.

\begin{figure}
\begin{center}
\includegraphics [width=12cm]{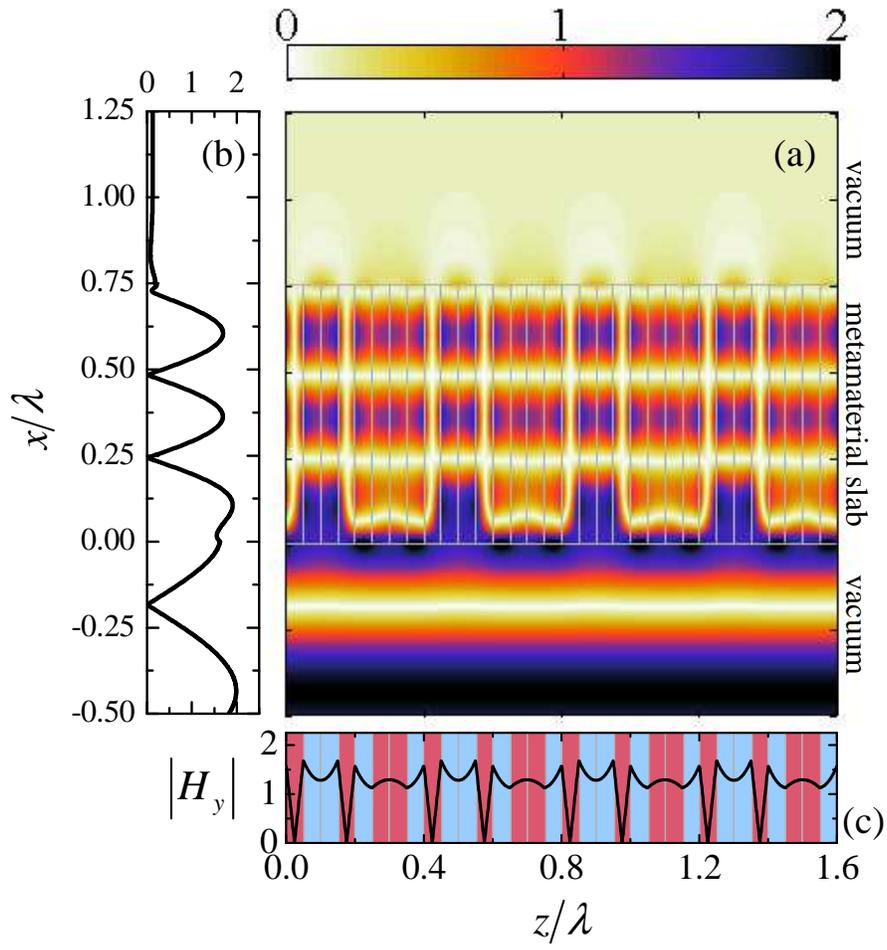}
\end{center}
\caption{(Color online) As in Fig. \ref{Figure3}, but for the $n=3$ iteration-order (four unit-cells shown). Longitudinal and transverse cuts in (b) and (c) are at $z=0.05\lambda$ and $x=0.375\lambda$, respectively.}
\label{Figure10}
\end{figure}

The response dramatically changes for the $n=3$ iteration order, as illustrated in Fig. \ref{Figure10}. In this case, a {\em standing-wave} pattern is clearly visible inside the slab. Looking at the longitudinal cut in Fig. \ref{Figure10}(b), from the distance of two consecutive peaks ($\sim 0.241\lambda$) we can estimate a propagation constant $k_x\approx 0.415 d/\pi$ which is in excellent agreement with the $k_{x0}=0.4138 d/\pi$ value pertaining to the degenerate additional extraordinary wave predicted by the EFCs in Fig. \ref{Figure2}(d) for $k_z=0$ (normal incidence). The $x-$cut in Fig. \ref{Figure10}(c), markedly different from the standard-periodic-multilayer counterparts in Figs. \ref{Figure8}(c) and \ref{Figure9}(c), shows a transverse field profile with much larger amplitude variations, and with peaks at the interfaces between positive- and negative-permittivity layers, thereby evidencing the {\em nonlocal} nature of this mode, due to the coupling of SPPs propagating along thee interfaces.

Qualitatively similar results can be observed for the higher-order iterations $n=4$ (Fig. \ref{Figure11}) and $n=5$ (Fig. \ref{Figure12}). However, the response in Fig. \ref{Figure12} is more complex than the previous cases, since the standing wave inside the slab now results from the interference of three waves (cf. Fig. \ref{Figure5}) with different wavenumbers. Also, a moderate increase (by a factor $\sim 3$) in the peak amplitudes is observed.

The above field maps validate our previous EFC-based studies, demonstrate the actual excitability of the additional extraordinary waves, and provide some useful insight into their physical nature. However, it would be interesting to relate these nonlocal effects to more practically accessible observables, and to investigate their sensitivity with respect to the geometrical parameters as well as the unavoidable material dispersion and losses. To this aim, for the same parameter configuration above, Fig. \ref{Figure13} compares the transmittances observed for the $n=1$ and $n=3$ iteration-orders, as a function of the slab thickness, at a fixed frequency.
While for the $n=1$ iteration-order (i.e., standard periodic case) the transmittance exhibits a very fast monotonic decay, for the $n=3$ iteration-order we can observe a series of sharp peaks characterized by almost perfect transmission. 
For this latter case, Fig. \ref{Figure14} shows a more realistic response, as a function of the normalized frequency. More specifically, we assume $\omega_0$ as a fiducial angular frequency, and consider a Drude-type dispersion model for the negative-permittivity material constituent,
\beq
\varepsilon_b\left(\omega\right)=1-\frac{\omega_p^2}{\omega\left(\omega+i\gamma\right)},
\label{eq:Drude}
\eeq
with parameters (given in the caption) chosen so that $\mbox{Re}[\varepsilon_{b}(\omega_0)]=-1.83$ with a loss-tangent of $10^{-3}$. We still observe a series of sharp transmittance peaks, though with a reduced dynamic range, due to the various detuning effects as well as the losses. 

As anticipated in Sec. \ref{Sec:NLP}, the nonlocal effective parameters in (\ref{eq:effNL}) are generally inadequate to providing an accurate description of the slab response. Nevertheless, they can still correctly predict some coarse features. For instance, by estimating via (\ref{eq:effNL}) the {\em effective material wavelength} (for normal incidence, i.e., $k_z=0$)
\beq
\lambda_e=\frac{\lambda}{\sqrt{{\hat \varepsilon}_{\perp}\left(0\right)}}
\label{eq:lambdae}
\eeq
as a function of frequency, we observe that the transmittance peaks in Fig. \ref{Figure14} occur at frequencies for which the slab thickness approaches a half-integer number of $\lambda_e$ (see the inset). Therefore, these peaks are attributable to Fabry-Perot-type resonances of the additional extraordinary waves.

\section{Some Remarks}
\label{Sec:Remarks}
\subsection{The Role of Aperiodic Order}
\label{Sec:Aperiodic}
Since the above results refer to various {\em finite} iteration-orders of the ThM geometry, one may wonder to what extent they are attributable to the inherent periodic truncation of the supercell (cf. Fig. \ref{Figure1}) or to more ore less trivial scaling effects, rather than the actual aperiodic order. The following considerations are in order. 

First, we highlight that the particular structure of the ThM inflation rule in (\ref{eq:SubRule}) ensures that, at any iteration-order, no more than two consecutive identical symbols may occur (e.g., $aaa$ and $bbb$, or longer, sequences are forbidden).\cite{Queffelec2010} This implies that the nonlocal effects observed are not trivially attributable to an effective increase of the average layer thickness.
 
Second, we recall that the trace-map formalism may be effectively utilized to infer some {\em asymptotic} properties in the limit $n\rightarrow\infty$, i.e., in the limit for which the artificial periodicity enforced by the Bloch-type phase-shift walls is washed out by the actual aperiodic order. Within this framework, we observe from Fig. \ref{Figure2} that the the stationary point at $k_{x0}$ corresponds to $\chi_3=2$, which represents the band-edge condition. When substituted in (\ref{eq:chi1pmap}) (together with ${\dot \chi}_3=0$), this implies that also ${\dot \chi_4}=0$ at $k_{x0}$. Moreover, we note from the trace-map (\ref{eq:tracemap}) that this will also imply that
\beq
\chi_n\left(k_{x0}^2\right)=2, ~~~ {\dot \chi_n}\left(k_{x0}^2\right)=0, ~~ n\ge 5.
\eeq
In other words, the additional extraordinary waves associated with the stationary point at $k_{x0}$ will be retained by any arbitrarily high iteration-order of the ThM multilayer, and hence also in the actual aperiodic-order limit.

\subsection{Relationship with Resonant Transmission}
\begin{figure}
\begin{center}
\includegraphics [width=12cm]{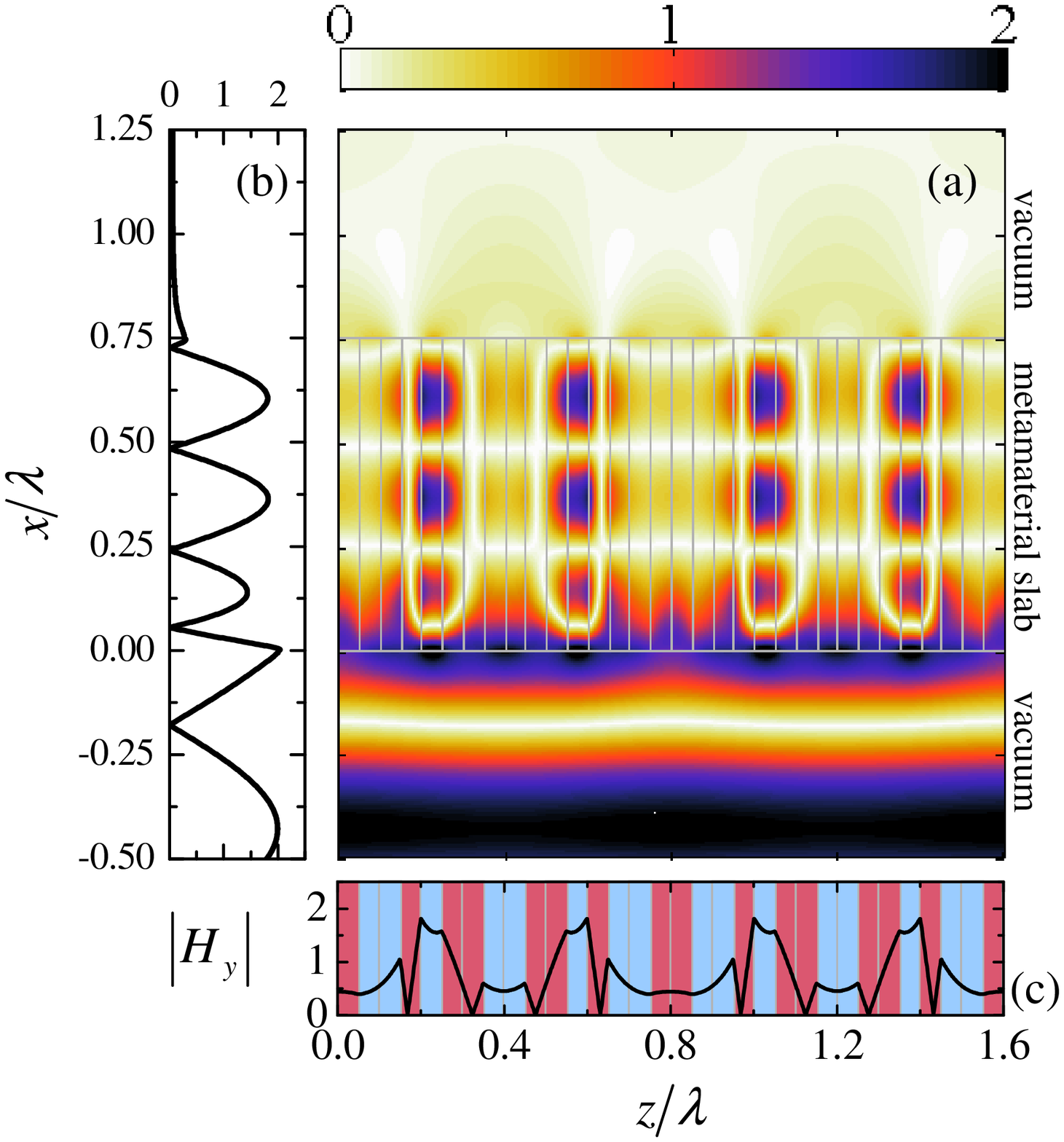}
\end{center}
\caption{(Color online) As in Fig. \ref{Figure3}, but for the $n=4$ iteration-order (two unit-cells shown). Cuts in (b) and (c) are at $z=0.2\lambda$ and $x=0.375\lambda$, respectively.}
\label{Figure11}
\end{figure}

\begin{figure}
\begin{center}
\includegraphics [width=12cm]{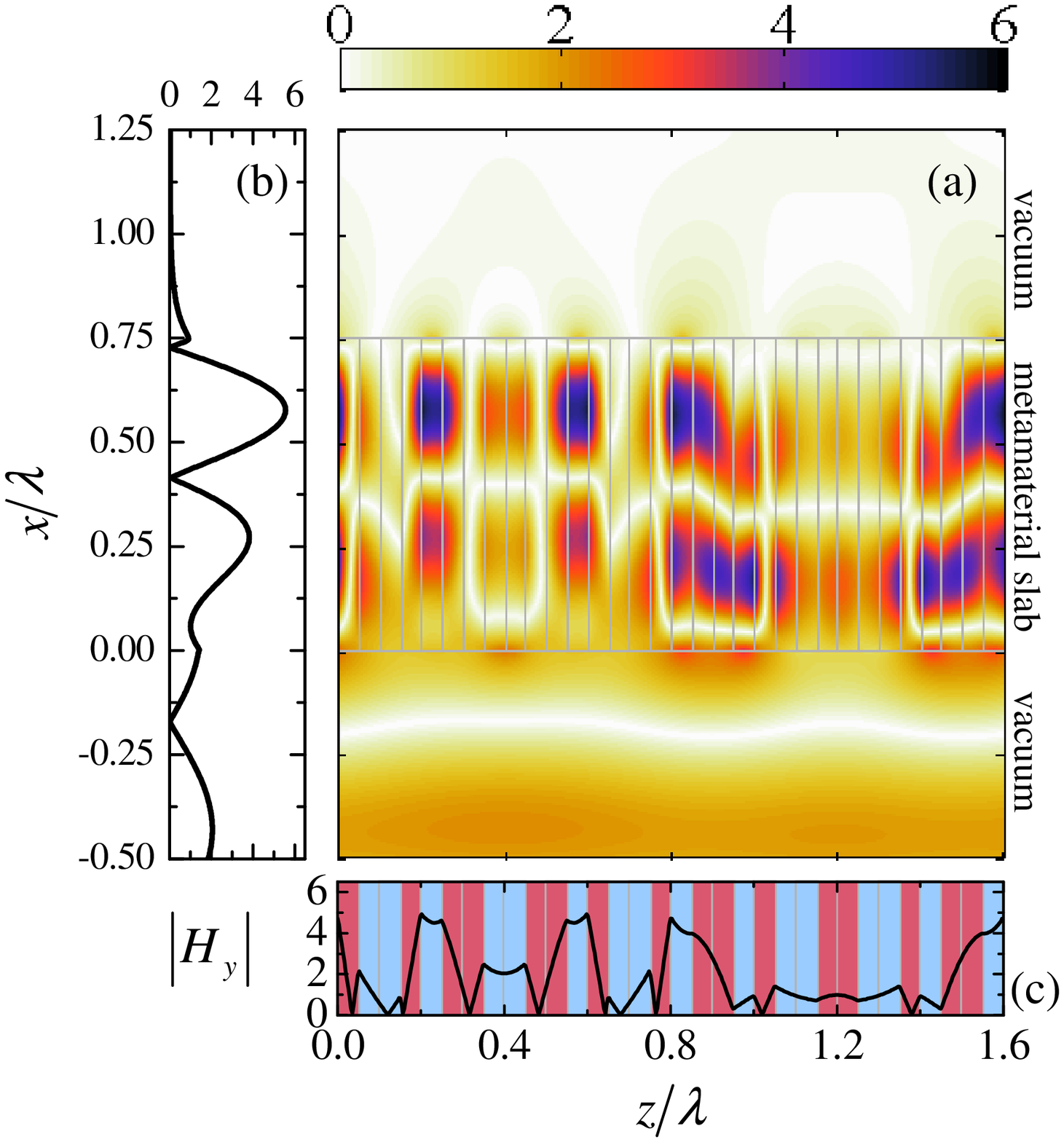}
\end{center}
\caption{(Color online) As in Fig. \ref{Figure3}, but for the $n=5$ iteration-order (one unit-cell shown). Cuts in (b) and (c) are at $z=0.2\lambda$ and $x=0.27\lambda$, respectively.}
\label{Figure12}
\end{figure}

\begin{figure}
\begin{center}
\includegraphics [width=12cm]{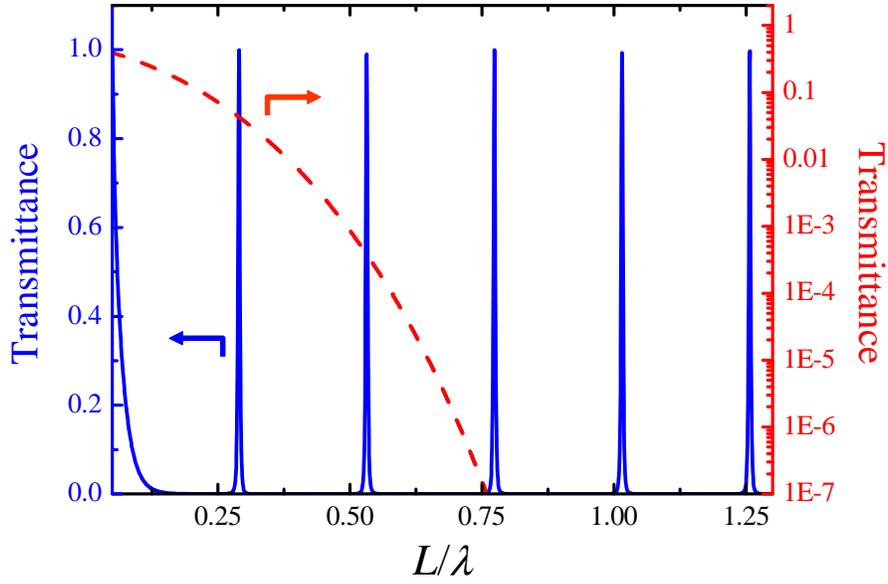}
\end{center}
\caption{(Color online) Geometry and parameters as in Figs. \ref{Figure8} and \ref{Figure10}. Transmittance as a function of the slab thickness, for the $n=1$ (red-dashed curve; right axis with logarithmic scale) and $n=3$ (blue-solid curve; left axis), at a fixed frequency.}
\label{Figure13}
\end{figure}

\begin{figure}
\begin{center}
\includegraphics [width=12cm]{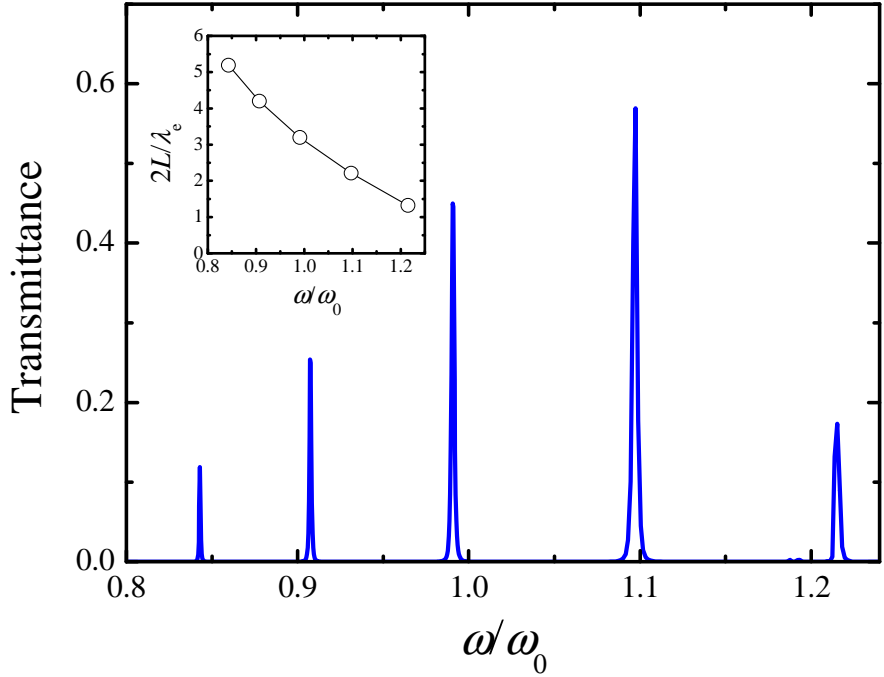}
\end{center}
\caption{(Color online) As in Fig. \ref{Figure13}, but for fixed slab thickness and the $n=3$ iteration-order, as a function of the normalized angular frequency. A fiducial angular frequency $\omega_0$ is assumed (with $\lambda_0$ denoting the corresponding wavelength), and the negative-permittivity constituent material is modeled via the Drude-type dispersion law in (\ref{eq:Drude}), with $\omega_p=1.682\omega_0$ and $\gamma=6.47\cdot10^{-4}\omega_0$, so that so that $\mbox{Re}[\varepsilon_{b}(\omega_0)]=-1.83$ with a loss-tangent of $10^{-3}$. The slab and constituent-layer thicknesses are chosen as $L=0.75\lambda_0$ and $d_a=d_b=d/2=0.05\lambda_0$. The inset shows the slab thickness normalized with respect to the effective material (half)wavelength in (\ref{eq:lambdae}).}
\label{Figure14}
\end{figure}

In previous works,\cite{Qiu2003,Hsueh2011} the condition $\chi_n=2$ has also been associated with perfect transmission through ThM-based quasicrystal multilayers. We emphasize that such configurations are, however, different from our slab configurations in Figs. \ref{Figure8}-\ref{Figure12}, since they assume a finite-size (along $z$) $n$th-order ThM dielectric multilayer sandwiched between two infinite (along $z$) homogeneous, isotropic media. Conversely, in our configurations in Figs. \ref{Figure8}-\ref{Figure12}, the slab is truncated along the $x$ direction, and the material constituents have oppositely-signed permittivities.

For the ThM multilayer configurations as in Refs. \onlinecite{Qiu2003,Hsueh2011}, It can be shown that the transmittance pertaining to a generic iteration-order $n$ can be parameterized as\cite{Wang2000}
\beq
T_n=\frac{4}{\chi_n^2+\upsilon_n^2},
\label{eq:Tn}
\eeq
with $\upsilon_n$ denoting the ``anti-trace'' (i.e., the difference between the off-diagonal terms $M_{21}$ and $M_{12}$) of the multilayer transfer-matrix ${\underline {\underline M}}$ (see Appendix \ref{Sec:APPA}). Interestingly, for the ThM sequence, also the anti-trace admits a simple map. In particular, letting $\upsilon_n$ and ${\tilde \upsilon}_n$ the transfer-matrix anti-traces pertaining to a ThM multilayer at iteration-order $n$ initiated by an ``$ab$'' or a ``$ba$'' sequence, respectively, the following recursive rules hold\cite{Wang2000}
\begin{eqnarray}
\upsilon_{n+1}&=&\chi_n\left[
\left(
\chi_n-1
\right)\upsilon_n+{\tilde \upsilon}_{n-1}
\right],\nonumber\\
{\tilde \upsilon}_{n+1}&=&\chi_n\left[
\left(
\chi_n-1
\right){\tilde \upsilon}_n+\upsilon_{n-1}
\right],~~~n\ge1.
\end{eqnarray}
In the lossless case (i.e., real-valued traces and anti-traces), we note that the condition $\chi_1=0$ (which ensures the presence of additional extraordinary waves in our scenario) also implies $\chi_{n}=2, {\upsilon}_n={\tilde \upsilon_n}=0$, i.e., $T_n=1$ for $n\ge 3$ in (\ref{eq:Tn}). This corresponds to perfect transmission in the scenarios of Refs. \onlinecite{Qiu2003,Hsueh2011}. Paralleling those studies, and generalizing  the underlying approaches (see, e.g., Ref. \onlinecite{Hsueh2011}), it can be shown  for our scenario that the number of additional extraordinary waves grows {\em exponentially} with the iteration-order. In particular, the spectrum exhibits self-similarity and characteristic trifurcation.\citep{Qiu2003}

\subsection{The Case $\varepsilon_{\parallel}>0$, $\varepsilon_{\perp}<0$}
Although all results above pertain to a parameter configuration featuring $\varepsilon_{\parallel}>0$ and $\varepsilon_{\perp}<0$, qualitatively similar considerations also hold for parameter configurations characterized by $\varepsilon_{\parallel}<0$ and $\varepsilon_{\perp}>0$.

\begin{figure}
\begin{center}
\includegraphics [width=12cm]{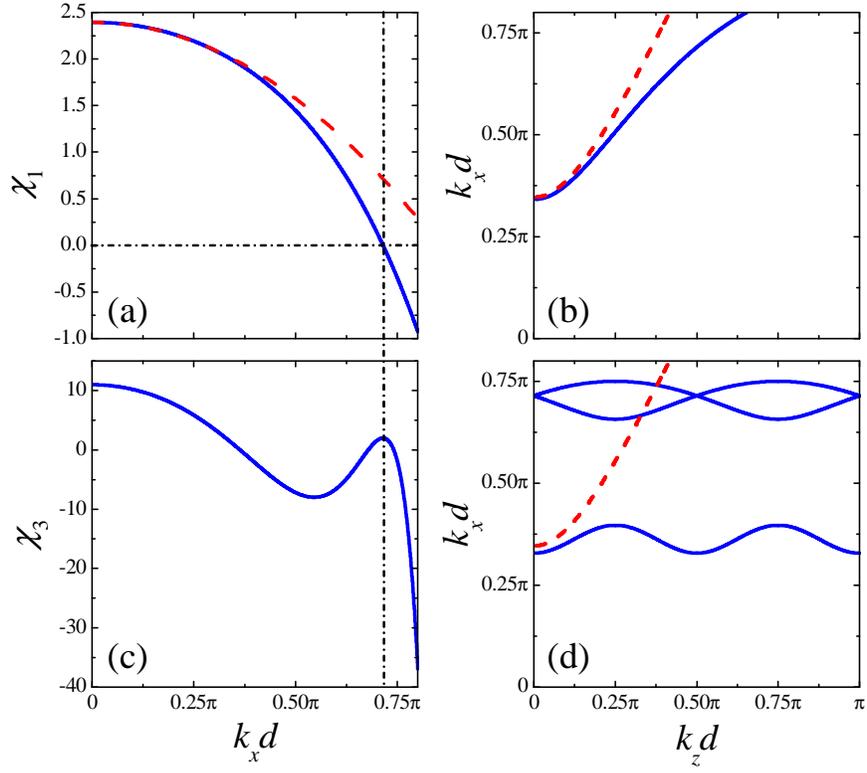}
\end{center}
\caption{(Color online) As in Fig. \ref{Figure2}, but for $\varepsilon_a=1$, $\varepsilon_b=-3$   (i.e, $\varepsilon_{\parallel}=-1$, $\varepsilon_{\perp}=3$).}
\label{Figure15}
\end{figure}

Figure \ref{Figure15} shows the traces and corresponding EFCs at iteration-orders $n=1$  and $n=2$, as in Fig. \ref{Figure2}, but with different material parameters (given in the caption) corresponding to $\varepsilon_{\parallel}=-1$  and $\varepsilon_{\perp}=3$. Once again, parameters are chosen so that the first iteration  $n=1$ (standard periodic multilayer) is reasonably well described by the local (EMT) model. As can be observed from Fig. \ref{Figure15}(a), the trace $\chi_1$ agrees pretty well with its quadratic approximation in (\ref{eq:chi1app}) up to values $k_x\lesssim 0.5\pi/d$, and vanishes at $k_{x0}=0.715\pi/d$. Therefore, by comparison with the case in Fig. \ref{Figure2}, the simple analytical approximation in (\ref{eq:kx0}) now yields a larger ($\sim 20\%$) error in the position of the zero. Nonetheless, as can be observed from Fig. \ref{Figure15}(b), the local EMT model still correctly predicts the presence of a single wave propagating for arbitrarily small values of $k_z$.
As expected, the trace $\chi_3$  pertaining to the $n=3$ iteration-order exhibits a maximum ($\chi_3=2$) at $k_{x0}$  [see Fig. \ref{Figure15}(c)]. However, unlike the case in Fig. \ref{Figure15}(c), it now also exhibits a local minimum at $k_x=0.546\pi/d$, which arises from the vanishing of the term in square brackets in (\ref{eq:chi1pmap}). This yields in the corresponding EFCs [Fig. \ref{Figure15}(d)] {\em three} (always propagating) modes, one of which approaches the local EMT prediction for small values of $k_z$. The other two instead represent additional extraordinary waves which degenerate at $k_z^{(m)}=m\pi/(2d)$, $m=0,1,2,...$  Therefore, the same general observations and conclusions hold as for the case featuring $\varepsilon_{\parallel}>0$ and $\varepsilon_{\perp}<0$ in Fig. \ref{Figure2}. However, the field distributions for plane-wave-excited slab configurations (not shown for brevity) are less clean-cut than those in Figs. \ref{Figure8} and \ref{Figure10}, since the local mode is now always propagating, and thus the difference between the standard periodic multilayer and the $n=3$ iteration-order is less striking.

\section{Conclusions}
\label{Sec:Conclusions}
To sum up, we have shown that hyperbolic metamaterials implemented as multilayered based on the ThM sequence may exhibit strong optical nonlocality, manifested as the appearance of additional extraordinary waves and a strong wavevector dependence in the effective constitutive parameters. From the mathematical viewpoint, we associated these effects to stationary points of the transfer-matrix trace, and derived simple analytical design rules. The chosen ThM geometry is particularly interesting since different iteration-orders differ solely in the positional order of the constituent material layers. Interestingly, we identified some configurations for which these nonlocal effects are rather weak at the first two iteration-orders ($n=1,2$, corresponding to standard periodic multilayers) and become markedly more prominent at higher iteration-orders $n\ge 3$, even in the limit $n\rightarrow\infty$ for which the (periodic) truncation effects become progressively less relevant.

To the best of our knowledge, against the many implications and applications of aperiodic-order to optics and photonics (see, e.g., Refs. \onlinecite{Macia2006,DalNegro2012} for recent reviews), this represents the first evidence in connection with optical nonlocality.
Besides the inherent academic interest, from the application viewpoint, this constitutes an important, and technologically inexpensive, additional degree of freedom in the engineering of optical nonlocality, which may be also be fruitfully exploited within the recently-introduced framework of nonlocal transformation optics.\cite{Castaldi2012}

We highlight that the ThM sequence was considered here only in view of its particularly simple inflation rule and associated trace-map, which facilitate analytical treatment as well as direct comparison with standard periodic multilayers, but the results are more general. In fact, one of the most intriguing follow-up study may be the systematic design of deterministic aperiodic sequences, via suitable inflation rules and associated polynomial trace-maps,\cite{Kolar1990} yielding prescribed nonlocal effects.

\appendix

\section{Dispersion Relationships for Generic Multilayers}
\label{Sec:APPA}
\begin{figure}
\begin{center}
\includegraphics [width=10cm]{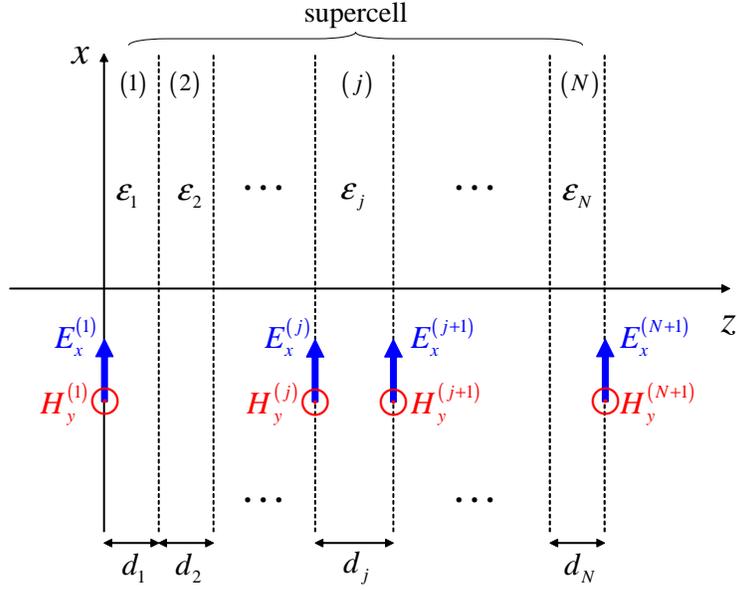}
\end{center}
\caption{(Color online) Geometry of a general supercell composed of  $N$ layers, with relative permittivities $\varepsilon_j$  and thicknesses $d_j$, $j=1,...,N$, stacked along the $z$-direction. Also shown are the tangential field components $E_x^{(j)}$  and $H_y^{(j)}$  at the layer interfaces, relevant for TM polarization.}
\label{Figure16}
\end{figure}

The dispersion relationship of a periodic multilayer consisting of the infinite replication of a generic supercell can be straightforwardly obtained by applying the rigorous transfer-matrix approach.\cite{Born1999} Figure \ref{Figure16} schematically illustrates a rather general supercell composed of $N$  layers with relative permittivity $\varepsilon_j$  and thickness $d_j$, $j=1,...,N$, stacked along the $z$-direction.
For the assumed TM polarization, the tangential components at the two interfaces of a generic  $j$-th layer can be related as
\beq
\left[
\begin{array}{cc}
E_x^{(j)}\\
i\eta H_y^{(j)}
\end{array}
\right]={\underline {\underline M}}^{(j)}\cdot \left[
\begin{array}{cc}
E_x^{(j+1)}\\
i\eta H_y^{(j+1)}
\end{array}
\right],
\label{eq:ExHy1}
\eeq
where $\eta$ denotes the vacuum characteristic impedance, the $(x,z)$ dependence in the fields has been omitted for notational compactness, and the {\em unimodular} transfer matrix is given by
\beq
{\underline {\underline M}}^{(j)}=
\left[
\begin{array}{cc}
\cos\delta_j & \displaystyle{-\frac{\sin\delta_j}{\gamma_j}}\\
\gamma_j\sin\delta_j & \cos\delta_j
\end{array}
\right],
\label{eq:TF}
\eeq
with
\beq
\delta_j=k_{zj}d_j,~~\gamma_j=\frac{\varepsilon_j k}{k_{zj}},~~k_{zj}=\sqrt{k^2\varepsilon_j-k_x^2},~~\mbox{Im}\left(k_{zj}\right)\ge0,
\label{eq:deltaj}
\eeq 
and $k_x$ indicating the (conserved) transverse wavenumber. Proceeding layer by layer, the tangential field components at the interfaces of the supercell can be therefore related by cascading the relevant transfer matrices, viz.,
\beq
\left[
\begin{array}{cc}
E_x^{(1)}\\
i\eta H_y^{(1)}
\end{array}
\right]={\underline {\underline M}}\cdot \left[
\begin{array}{cc}
E_x^{(N+1)}\\
i\eta H_y^{(N+1)}
\end{array}
\right],
\eeq
with
\beq
{\underline {\underline M}}=\prod_{j=1}^N{\underline {\underline M}}^{(j)}=\left[
\begin{array}{cc}
M_{11} & M_{12}\\
M_{21} & M_{22}
\end{array}
\right].
\label{eq:MMat}
\eeq
In view of the assumed overall periodicity along $z$, the fields should be likewise periodic of period $L$  (the overall thickness of the supercell), apart from a phase factor. This can be enforced via Bloch-type phase-shift conditions at the supercell interfaces $z=0$ and $z=L$
\beq
\left[
\begin{array}{cc}
E_x^{(1)}\\
i\eta H_y^{(1)}
\end{array}
\right]=\exp\left(-ik_{zB}L\right) \left[
\begin{array}{cc}
E_x^{(N+1)}\\
i\eta H_y^{(N+1)}
\end{array}
\right],
\label{eq:Bloch1}
\eeq
with $k_{zB}$ denoting the Bloch propagation constant. The arising homogeneous linear system of equations admit nontrivial solutions if
\beq
\det\left[
{\underline {\underline M}}-\exp\left(-ik_{zB}L\right){\underline {\underline I}}
\right]=0,
\eeq
with ${\underline {\underline I}}$ denoting the $2\times 2$ identity matrix. Recalling that the supercell transfer-matrix ${\underline {\underline M}}$  in (\ref{eq:MMat}) is unimodular (as the product of unimodular matrices), i.e.
\beq
\det\left({\underline {\underline M}}\right)=M_{11}M_{22}-M_{12}M_{21}=1,
\eeq
the relationship in (\ref{eq:Bloch1}) can be recast as
\beq
\cos\left(k_{zB}L\right)=\frac{1}{2}\left(M_{11}+M_{22}\right)=\frac{1}{2}\mbox{Tr}\left({\underline {\underline M}}\right),
\eeq
with $\mbox{Tr}$ denoting the {\em trace} operator.\citep{Lang1987}

\section{Trace Map for Thue-Morse Superlattices}
\label{Sec:APPB}
For a multilayer composed of only two types of layers (labeled as ``$a$'' and ``$b$'') arranged according to the ThM aperiodic sequence, of interest in our study, the trace of the supercell transfer-matrix at a generic iteration-order $n$ needs not to be calculated via the product in (\ref{eq:MMat}), but can be obtained in a much more direct fashion, via the trace-map formalism.\cite{Kolar1990,Wang2000} Letting ${\underline {\underline M}}_n$ and ${\tilde {\underline {\underline M}}}_n$ the transfer matrices pertaining to a ThM multilayer at iteration-order $n$ initiated by an ``$ab$''-type or a ``$ba$''-type sequence, respectively, the following recursive rules hold \cite{Grigoriev2010}
\beq
{\underline {\underline M}}_{n+1}={\underline {\underline M}}_n\cdot {\tilde {\underline {\underline M}}}_n,~~~
{\tilde {\underline {\underline M}}}_{n+1}={\tilde {\underline {\underline M}}}_n\cdot {\underline {\underline M}}_n,
\eeq
with initial conditions
\beq
{\underline {\underline M}}_0={\underline {\underline M}}^{(a)},~~~{\tilde {\underline {\underline M}}}_0={\underline {\underline M}}^{(b)},
\eeq
where ${\underline {\underline M}}^{(a)}$ and ${\underline {\underline M}}^{(b)}$ denote the transfer matrices associated with a single $a$-type and $b$-type layer, respectively. These latter may be formally obtained via (\ref{eq:ExHy1})--(\ref{eq:deltaj}) by replacing the layer index $j$ with the symbols $a$ and $b$, respectively. Thus, for a given iteration-order $n\ge 1$, we obtain
\begin{eqnarray}
\chi_n&\equiv&\mbox{Tr}\left({\underline {\underline M}}_n\right)=\mbox{Tr}\left({\underline {\underline M}}_{n-1}\cdot {\tilde {\underline {\underline M}}}_{n-1}\right)\nonumber\\
&=&\mbox{Tr}\left({\tilde {\underline {\underline M}}}_{n-1}\cdot{\underline {\underline M}}_{n-1}\right)=\mbox{Tr}\left({\tilde {\underline {\underline M}}}_n\right),
\label{eq:Tr0}
\end{eqnarray}
while, for adjacent iteration-orders, we can write
\begin{eqnarray}
\chi_{n+2}&=&\mbox{Tr}\left({\underline {\underline M}}_{n+2}\right)=\mbox{Tr}\left({\underline {\underline M}}_{n+1}\cdot {\tilde {\underline {\underline M}}}_{n+1}\right)\nonumber\\
&=&\mbox{Tr}\left({\underline {\underline M}}_{n}\cdot {\tilde {\underline {\underline M}}}_{n} \cdot {\tilde {\underline {\underline M}}}_{n}\cdot {\underline {\underline M}}_{n}\right)\nonumber\\
&=&\mbox{Tr}\left[\left({\underline {\underline M}}_n\right)^2\cdot\left({\tilde {\underline {\underline M}}}_n\right)^2\right],~~n\ge1,
\label{eq:Tr1}
\end{eqnarray}
with the last equality following from the trace invariance under cyclic permutations.\citep{Lang1987}
Recalling that, as a consequence of the Caley-Hamilton theorem,\cite{Lang1987} the square of a general unimodular matrix ${\underline {\underline A}}$ of trace $\chi_A$ can be written as
\beq
{\underline {\underline A}}^2=\chi_A{\underline {\underline A}}-{\underline {\underline I}},
\eeq
and substituting in (\ref{eq:Tr1}), we obtain
\begin{eqnarray}
\chi_{n+2}&=&\mbox{Tr}\left[
\left(\chi_n{\underline {\underline M}}_n-{\underline {\underline I}}\right)
\cdot
\left(\chi_n{\tilde {\underline {\underline M}}}_n-{\underline {\underline I}}\right)
\right]\nonumber\\
&=&
\mbox{Tr}\left(
\chi_n^2{\underline {\underline M}}_n\cdot {\tilde {\underline {\underline M}}}_n- \chi_n{\underline {\underline M}}_n -\chi_n {\tilde {\underline {\underline M}}}_n+{\tilde {\underline {\underline I}}}
\right)\nonumber\\
&=&\chi_n^2\left(\chi_{n+1}-2\right)+2,~~~n\ge1,
\end{eqnarray}
where, in the last equality, the linearity of the trace operator has been exploited, as well as the result in (\ref{eq:Tr0}). This yields the trace-map in (\ref{eq:tracemap}).


\begin{thebibliography}{36}%
\makeatletter
\providecommand \@ifxundefined [1]{%
 \@ifx{#1\undefined}
}%
\providecommand \@ifnum [1]{%
 \ifnum #1\expandafter \@firstoftwo
 \else \expandafter \@secondoftwo
 \fi
}%
\providecommand \@ifx [1]{%
 \ifx #1\expandafter \@firstoftwo
 \else \expandafter \@secondoftwo
 \fi
}%
\providecommand \natexlab [1]{#1}%
\providecommand \enquote  [1]{``#1''}%
\providecommand \bibnamefont  [1]{#1}%
\providecommand \bibfnamefont [1]{#1}%
\providecommand \citenamefont [1]{#1}%
\providecommand \href@noop [0]{\@secondoftwo}%
\providecommand \href [0]{\begingroup \@sanitize@url \@href}%
\providecommand \@href[1]{\@@startlink{#1}\@@href}%
\providecommand \@@href[1]{\endgroup#1\@@endlink}%
\providecommand \@sanitize@url [0]{\catcode `\\12\catcode `\$12\catcode
  `\&12\catcode `\#12\catcode `\^12\catcode `\_12\catcode `\%12\relax}%
\providecommand \@@startlink[1]{}%
\providecommand \@@endlink[0]{}%
\providecommand \url  [0]{\begingroup\@sanitize@url \@url }%
\providecommand \@url [1]{\endgroup\@href {#1}{\urlprefix }}%
\providecommand \urlprefix  [0]{URL }%
\providecommand \Eprint [0]{\href }%
\providecommand \doibase [0]{http://dx.doi.org/}%
\providecommand \selectlanguage [0]{\@gobble}%
\providecommand \bibinfo  [0]{\@secondoftwo}%
\providecommand \bibfield  [0]{\@secondoftwo}%
\providecommand \translation [1]{[#1]}%
\providecommand \BibitemOpen [0]{}%
\providecommand \bibitemStop [0]{}%
\providecommand \bibitemNoStop [0]{.\EOS\space}%
\providecommand \EOS [0]{\spacefactor3000\relax}%
\providecommand \BibitemShut  [1]{\csname bibitem#1\endcsname}%
\let\auto@bib@innerbib\@empty
\bibitem [{\citenamefont {Capolino}(2009)}]{Capolino2009}%
  \BibitemOpen
  \bibfield  {author} {\bibinfo {author} {\bibfnamefont {F.}~\bibnamefont
  {Capolino}},\ }\href {http://books.google.it/books?id=0PMnYo8hva8C} {\emph
  {\bibinfo {title} {Metamaterials Handbook}}},\ \bibinfo {series}
  {Metamaterials Handbook}, Vol.\ \bibinfo {volume} {1 and 2}\ (\bibinfo
  {publisher} {CRC Press, Boca Raton, FL, USA},\ \bibinfo {year}
  {2009})\BibitemShut {NoStop}%
\bibitem [{\citenamefont {Smith}\ \emph {et~al.}(2004)\citenamefont {Smith},
  \citenamefont {Kolinko},\ and\ \citenamefont {Schurig}}]{Smith2004}%
  \BibitemOpen
  \bibfield  {author} {\bibinfo {author} {\bibfnamefont {D.~R.}\ \bibnamefont
  {Smith}}, \bibinfo {author} {\bibfnamefont {P.}~\bibnamefont {Kolinko}}, \
  and\ \bibinfo {author} {\bibfnamefont {D.}~\bibnamefont {Schurig}},\ }\href
  {\doibase 10.1364/JOSAB.21.001032} {\bibfield  {journal} {\bibinfo  {journal}
  {J. Opt. Soc. Am. B}\ }\textbf {\bibinfo {volume} {21}},\ \bibinfo {pages}
  {1032} (\bibinfo {year} {2004})}\BibitemShut {NoStop}%
\bibitem [{\citenamefont {Noginov}\ \emph {et~al.}(2009)\citenamefont
  {Noginov}, \citenamefont {Barnakov}, \citenamefont {Zhu}, \citenamefont
  {Tumkur}, \citenamefont {Li},\ and\ \citenamefont {Narimanov}}]{Noginov2009}%
  \BibitemOpen
  \bibfield  {author} {\bibinfo {author} {\bibfnamefont {M.~A.}\ \bibnamefont
  {Noginov}}, \bibinfo {author} {\bibfnamefont {Y.~A.}\ \bibnamefont
  {Barnakov}}, \bibinfo {author} {\bibfnamefont {G.}~\bibnamefont {Zhu}},
  \bibinfo {author} {\bibfnamefont {T.}~\bibnamefont {Tumkur}}, \bibinfo
  {author} {\bibfnamefont {H.}~\bibnamefont {Li}}, \ and\ \bibinfo {author}
  {\bibfnamefont {E.~E.}\ \bibnamefont {Narimanov}},\ }\href {\doibase
  10.1063/1.3115145} {\bibfield  {journal} {\bibinfo  {journal} {Appl. Phys.
  Lett.}\ }\textbf {\bibinfo {volume} {94}},\ \bibinfo {eid} {151105} (\bibinfo
  {year} {2009})}\BibitemShut {NoStop}%
\bibitem [{\citenamefont {Jacob}\ \emph {et~al.}(2006)\citenamefont {Jacob},
  \citenamefont {Alekseyev},\ and\ \citenamefont {Narimanov}}]{Jacob2006}%
  \BibitemOpen
  \bibfield  {author} {\bibinfo {author} {\bibfnamefont {Z.}~\bibnamefont
  {Jacob}}, \bibinfo {author} {\bibfnamefont {L.~V.}\ \bibnamefont
  {Alekseyev}}, \ and\ \bibinfo {author} {\bibfnamefont {E.}~\bibnamefont
  {Narimanov}},\ }\href {\doibase 10.1364/OE.14.008247} {\bibfield  {journal}
  {\bibinfo  {journal} {Opt. Express}\ }\textbf {\bibinfo {volume} {14}},\
  \bibinfo {pages} {8247} (\bibinfo {year} {2006})}\BibitemShut {NoStop}%
\bibitem [{\citenamefont {Govyadinov}\ and\ \citenamefont
  {Podolskiy}(2006)}]{Govyadinov2006}%
  \BibitemOpen
  \bibfield  {author} {\bibinfo {author} {\bibfnamefont {A.~A.}\ \bibnamefont
  {Govyadinov}}\ and\ \bibinfo {author} {\bibfnamefont {V.~A.}\ \bibnamefont
  {Podolskiy}},\ }\href {\doibase 10.1103/PhysRevB.73.155108} {\bibfield
  {journal} {\bibinfo  {journal} {Phys. Rev. B}\ }\textbf {\bibinfo {volume}
  {73}},\ \bibinfo {pages} {155108} (\bibinfo {year} {2006})}\BibitemShut
  {NoStop}%
\bibitem [{\citenamefont {Smolyaninov}\ and\ \citenamefont
  {Narimanov}(2010)}]{Smolyaninov2010}%
  \BibitemOpen
  \bibfield  {author} {\bibinfo {author} {\bibfnamefont {I.~I.}\ \bibnamefont
  {Smolyaninov}}\ and\ \bibinfo {author} {\bibfnamefont {E.~E.}\ \bibnamefont
  {Narimanov}},\ }\href {\doibase 10.1103/PhysRevLett.105.067402} {\bibfield
  {journal} {\bibinfo  {journal} {Phys. Rev. Lett.}\ }\textbf {\bibinfo
  {volume} {105}},\ \bibinfo {pages} {067402} (\bibinfo {year}
  {2010})}\BibitemShut {NoStop}%
\bibitem [{\citenamefont {Jacob}\ \emph {et~al.}(2010)\citenamefont {Jacob},
  \citenamefont {Kim}, \citenamefont {Naik}, \citenamefont {Boltasseva},
  \citenamefont {Narimanov},\ and\ \citenamefont {Shalaev}}]{Jacob2010}%
  \BibitemOpen
  \bibfield  {author} {\bibinfo {author} {\bibfnamefont {Z.}~\bibnamefont
  {Jacob}}, \bibinfo {author} {\bibfnamefont {J.-Y.}\ \bibnamefont {Kim}},
  \bibinfo {author} {\bibfnamefont {G.}~\bibnamefont {Naik}}, \bibinfo {author}
  {\bibfnamefont {A.}~\bibnamefont {Boltasseva}}, \bibinfo {author}
  {\bibfnamefont {E.}~\bibnamefont {Narimanov}}, \ and\ \bibinfo {author}
  {\bibfnamefont {V.}~\bibnamefont {Shalaev}},\ }\href {\doibase
  10.1007/s00340-010-4096-5} {\bibfield  {journal} {\bibinfo  {journal} {Appl.
  Phys. B}\ }\textbf {\bibinfo {volume} {100}},\ \bibinfo {pages} {215}
  (\bibinfo {year} {2010})}\BibitemShut {NoStop}%
\bibitem [{\citenamefont {Yao}\ \emph {et~al.}(2011)\citenamefont {Yao},
  \citenamefont {Yang}, \citenamefont {Yin}, \citenamefont {Bartal},\ and\
  \citenamefont {Zhang}}]{Yao2011}%
  \BibitemOpen
  \bibfield  {author} {\bibinfo {author} {\bibfnamefont {J.}~\bibnamefont
  {Yao}}, \bibinfo {author} {\bibfnamefont {X.}~\bibnamefont {Yang}}, \bibinfo
  {author} {\bibfnamefont {X.}~\bibnamefont {Yin}}, \bibinfo {author}
  {\bibfnamefont {G.}~\bibnamefont {Bartal}}, \ and\ \bibinfo {author}
  {\bibfnamefont {X.}~\bibnamefont {Zhang}},\ }\href@noop {} {\bibfield
  {journal} {\bibinfo  {journal} {Proc. Natl. Acad. Sci.}\ }\textbf {\bibinfo
  {volume} {108}},\ \bibinfo {pages} {11327} (\bibinfo {year}
  {2011})}\BibitemShut {NoStop}%
\bibitem [{\citenamefont {Krishnamoorthy}\ \emph {et~al.}(2012)\citenamefont
  {Krishnamoorthy}, \citenamefont {Jacob}, \citenamefont {Narimanov},
  \citenamefont {Kretzschmar},\ and\ \citenamefont
  {Menon}}]{Krishnamoorthy2012}%
  \BibitemOpen
  \bibfield  {author} {\bibinfo {author} {\bibfnamefont {H.~N.~S.}\
  \bibnamefont {Krishnamoorthy}}, \bibinfo {author} {\bibfnamefont
  {Z.}~\bibnamefont {Jacob}}, \bibinfo {author} {\bibfnamefont
  {E.}~\bibnamefont {Narimanov}}, \bibinfo {author} {\bibfnamefont
  {I.}~\bibnamefont {Kretzschmar}}, \ and\ \bibinfo {author} {\bibfnamefont
  {V.~M.}\ \bibnamefont {Menon}},\ }\href {\doibase 10.1126/science.1219171}
  {\bibfield  {journal} {\bibinfo  {journal} {Science}\ }\textbf {\bibinfo
  {volume} {336}},\ \bibinfo {pages} {205} (\bibinfo {year}
  {2012})}\BibitemShut {NoStop}%
\bibitem [{\citenamefont {Jacob}\ \emph {et~al.}(2012)\citenamefont {Jacob},
  \citenamefont {Smolyaninov},\ and\ \citenamefont {Narimanov}}]{Jacob2012}%
  \BibitemOpen
  \bibfield  {author} {\bibinfo {author} {\bibfnamefont {Z.}~\bibnamefont
  {Jacob}}, \bibinfo {author} {\bibfnamefont {I.~I.}\ \bibnamefont
  {Smolyaninov}}, \ and\ \bibinfo {author} {\bibfnamefont {E.~E.}\ \bibnamefont
  {Narimanov}},\ }\href@noop {} {\bibfield  {journal} {\bibinfo  {journal}
  {Appl. Phys. Lett.}\ }\textbf {\bibinfo {volume} {100}},\ \bibinfo {eid}
  {181105} (\bibinfo {year} {2012})}\BibitemShut {NoStop}%
\bibitem [{\citenamefont {Cortes}\ \emph {et~al.}(2012)\citenamefont {Cortes},
  \citenamefont {Newman}, \citenamefont {Molesky},\ and\ \citenamefont
  {Jacob}}]{Cortes2012}%
  \BibitemOpen
  \bibfield  {author} {\bibinfo {author} {\bibfnamefont {C.~L.}\ \bibnamefont
  {Cortes}}, \bibinfo {author} {\bibfnamefont {W.}~\bibnamefont {Newman}},
  \bibinfo {author} {\bibfnamefont {S.}~\bibnamefont {Molesky}}, \ and\
  \bibinfo {author} {\bibfnamefont {Z.}~\bibnamefont {Jacob}},\ }\href
  {http://stacks.iop.org/2040-8986/14/i=6/a=063001} {\bibfield  {journal}
  {\bibinfo  {journal} {J. Opt.}\ }\textbf {\bibinfo {volume} {14}},\ \bibinfo
  {pages} {063001} (\bibinfo {year} {2012})}\BibitemShut {NoStop}%
\bibitem [{\citenamefont {Biehs}\ \emph {et~al.}(2012)\citenamefont {Biehs},
  \citenamefont {Tschikin},\ and\ \citenamefont {Ben-Abdallah}}]{Biehs2012}%
  \BibitemOpen
  \bibfield  {author} {\bibinfo {author} {\bibfnamefont {S.-A.}\ \bibnamefont
  {Biehs}}, \bibinfo {author} {\bibfnamefont {M.}~\bibnamefont {Tschikin}}, \
  and\ \bibinfo {author} {\bibfnamefont {P.}~\bibnamefont {Ben-Abdallah}},\
  }\href {\doibase 10.1103/PhysRevLett.109.104301} {\bibfield  {journal}
  {\bibinfo  {journal} {Phys. Rev. Lett.}\ }\textbf {\bibinfo {volume} {109}},\
  \bibinfo {pages} {104301} (\bibinfo {year} {2012})}\BibitemShut {NoStop}%
\bibitem [{\citenamefont {Sihvola}(1999)}]{Sihvola1999}%
  \BibitemOpen
  \bibfield  {author} {\bibinfo {author} {\bibfnamefont {A.}~\bibnamefont
  {Sihvola}},\ }\href {http://books.google.it/books?id=uIHSNwxBxjgC} {\emph
  {\bibinfo {title} {Electromagnetic Mixing Formulas and Applications}}}\
  (\bibinfo  {publisher} {IEE Publishing, London},\ \bibinfo {year}
  {1999})\BibitemShut {NoStop}%
\bibitem [{\citenamefont {Elser}\ \emph {et~al.}(2007)\citenamefont {Elser},
  \citenamefont {Podolskiy}, \citenamefont {Salakhutdinov},\ and\ \citenamefont
  {Avrutsky}}]{Elser2007}%
  \BibitemOpen
  \bibfield  {author} {\bibinfo {author} {\bibfnamefont {J.}~\bibnamefont
  {Elser}}, \bibinfo {author} {\bibfnamefont {V.~A.}\ \bibnamefont
  {Podolskiy}}, \bibinfo {author} {\bibfnamefont {I.}~\bibnamefont
  {Salakhutdinov}}, \ and\ \bibinfo {author} {\bibfnamefont {I.}~\bibnamefont
  {Avrutsky}},\ }\href {\doibase 10.1063/1.2737935} {\bibfield  {journal}
  {\bibinfo  {journal} {Appl. Phys. Lett.}\ }\textbf {\bibinfo {volume} {90}},\
  \bibinfo {eid} {191109} (\bibinfo {year} {2007})}\BibitemShut {NoStop}%
\bibitem [{\citenamefont {Orlov}\ \emph {et~al.}(2011)\citenamefont {Orlov},
  \citenamefont {Voroshilov}, \citenamefont {Belov},\ and\ \citenamefont
  {Kivshar}}]{Orlov2011}%
  \BibitemOpen
  \bibfield  {author} {\bibinfo {author} {\bibfnamefont {A.~A.}\ \bibnamefont
  {Orlov}}, \bibinfo {author} {\bibfnamefont {P.~M.}\ \bibnamefont
  {Voroshilov}}, \bibinfo {author} {\bibfnamefont {P.~A.}\ \bibnamefont
  {Belov}}, \ and\ \bibinfo {author} {\bibfnamefont {Y.~S.}\ \bibnamefont
  {Kivshar}},\ }\href {\doibase 10.1103/PhysRevB.84.045424} {\bibfield
  {journal} {\bibinfo  {journal} {Phys. Rev. B}\ }\textbf {\bibinfo {volume}
  {84}},\ \bibinfo {pages} {045424} (\bibinfo {year} {2011})}\BibitemShut
  {NoStop}%
\bibitem [{\citenamefont {Chebykin}\ \emph {et~al.}(2012)\citenamefont
  {Chebykin}, \citenamefont {Orlov}, \citenamefont {Simovski}, \citenamefont
  {Kivshar},\ and\ \citenamefont {Belov}}]{Chebykin2012}%
  \BibitemOpen
  \bibfield  {author} {\bibinfo {author} {\bibfnamefont {A.~V.}\ \bibnamefont
  {Chebykin}}, \bibinfo {author} {\bibfnamefont {A.~A.}\ \bibnamefont {Orlov}},
  \bibinfo {author} {\bibfnamefont {C.~R.}\ \bibnamefont {Simovski}}, \bibinfo
  {author} {\bibfnamefont {Y.~S.}\ \bibnamefont {Kivshar}}, \ and\ \bibinfo
  {author} {\bibfnamefont {P.~A.}\ \bibnamefont {Belov}},\ }\href {\doibase
  10.1103/PhysRevB.86.115420} {\bibfield  {journal} {\bibinfo  {journal} {Phys.
  Rev. B}\ }\textbf {\bibinfo {volume} {86}},\ \bibinfo {pages} {115420}
  (\bibinfo {year} {2012})}\BibitemShut {NoStop}%
\bibitem [{\citenamefont {Kidwai}\ \emph {et~al.}(2012)\citenamefont {Kidwai},
  \citenamefont {Zhukovsky},\ and\ \citenamefont {Sipe}}]{Kidwai2012}%
  \BibitemOpen
  \bibfield  {author} {\bibinfo {author} {\bibfnamefont {O.}~\bibnamefont
  {Kidwai}}, \bibinfo {author} {\bibfnamefont {S.~V.}\ \bibnamefont
  {Zhukovsky}}, \ and\ \bibinfo {author} {\bibfnamefont {J.~E.}\ \bibnamefont
  {Sipe}},\ }\href {\doibase 10.1103/PhysRevA.85.053842} {\bibfield  {journal}
  {\bibinfo  {journal} {Phys. Rev. A}\ }\textbf {\bibinfo {volume} {85}},\
  \bibinfo {pages} {053842} (\bibinfo {year} {2012})}\BibitemShut {NoStop}%
\bibitem [{\citenamefont {Orlov}\ \emph {et~al.}(2013)\citenamefont {Orlov},
  \citenamefont {Iorsh}, \citenamefont {Belov},\ and\ \citenamefont
  {Kivshar}}]{Orlov2013}%
  \BibitemOpen
  \bibfield  {author} {\bibinfo {author} {\bibfnamefont {A.}~\bibnamefont
  {Orlov}}, \bibinfo {author} {\bibfnamefont {I.}~\bibnamefont {Iorsh}},
  \bibinfo {author} {\bibfnamefont {P.}~\bibnamefont {Belov}}, \ and\ \bibinfo
  {author} {\bibfnamefont {Y.}~\bibnamefont {Kivshar}},\ }\href@noop {}
  {\bibfield  {journal} {\bibinfo  {journal} {Opt. Express}\ }\textbf {\bibinfo
  {volume} {21}},\ \bibinfo {pages} {1593} (\bibinfo {year}
  {2013})}\BibitemShut {NoStop}%
\bibitem [{\citenamefont {Shechtman}\ \emph {et~al.}(1984)\citenamefont
  {Shechtman}, \citenamefont {Blech}, \citenamefont {Gratias},\ and\
  \citenamefont {Cahn}}]{Shechtman1984}%
  \BibitemOpen
  \bibfield  {author} {\bibinfo {author} {\bibfnamefont {D.}~\bibnamefont
  {Shechtman}}, \bibinfo {author} {\bibfnamefont {I.}~\bibnamefont {Blech}},
  \bibinfo {author} {\bibfnamefont {D.}~\bibnamefont {Gratias}}, \ and\
  \bibinfo {author} {\bibfnamefont {J.~W.}\ \bibnamefont {Cahn}},\ }\href
  {\doibase 10.1103/PhysRevLett.53.1951} {\bibfield  {journal} {\bibinfo
  {journal} {Phys. Rev. Lett.}\ }\textbf {\bibinfo {volume} {53}},\ \bibinfo
  {pages} {1951} (\bibinfo {year} {1984})}\BibitemShut {NoStop}%
\bibitem [{\citenamefont {Levine}\ and\ \citenamefont
  {Steinhardt}(1984)}]{Levine1984}%
  \BibitemOpen
  \bibfield  {author} {\bibinfo {author} {\bibfnamefont {D.}~\bibnamefont
  {Levine}}\ and\ \bibinfo {author} {\bibfnamefont {P.~J.}\ \bibnamefont
  {Steinhardt}},\ }\href {\doibase 10.1103/PhysRevLett.53.2477} {\bibfield
  {journal} {\bibinfo  {journal} {Phys. Rev. Lett.}\ }\textbf {\bibinfo
  {volume} {53}},\ \bibinfo {pages} {2477} (\bibinfo {year}
  {1984})}\BibitemShut {NoStop}%
\bibitem [{\citenamefont {Queff\'elec}(2010)}]{Queffelec2010}%
  \BibitemOpen
  \bibfield  {author} {\bibinfo {author} {\bibfnamefont {M.}~\bibnamefont
  {Queff\'elec}},\ }\href {http://books.google.it/books?id=V3gGuZwUIMAC} {\emph
  {\bibinfo {title} {Substitution Dynamical Systems -- Spectral Analysis}}},\
  Lecture Notes in Mathematics\ (\bibinfo  {publisher} {Springer-Verlag,
  Berlin-Heidelberg, Germany},\ \bibinfo {year} {2010})\BibitemShut {NoStop}%
\bibitem [{\citenamefont {Liu}(1997)}]{Liu1997}%
  \BibitemOpen
  \bibfield  {author} {\bibinfo {author} {\bibfnamefont {N.-H.}\ \bibnamefont
  {Liu}},\ }\href {\doibase 10.1103/PhysRevB.55.3543} {\bibfield  {journal}
  {\bibinfo  {journal} {Phys. Rev. B}\ }\textbf {\bibinfo {volume} {55}},\
  \bibinfo {pages} {3543} (\bibinfo {year} {1997})}\BibitemShut {NoStop}%
\bibitem [{\citenamefont {Qiu}\ \emph {et~al.}(2003)\citenamefont {Qiu},
  \citenamefont {Peng}, \citenamefont {Huang}, \citenamefont {Liu},
  \citenamefont {Wang}, \citenamefont {Hu},\ and\ \citenamefont
  {Jiang}}]{Qiu2003}%
  \BibitemOpen
  \bibfield  {author} {\bibinfo {author} {\bibfnamefont {F.}~\bibnamefont
  {Qiu}}, \bibinfo {author} {\bibfnamefont {R.~W.}\ \bibnamefont {Peng}},
  \bibinfo {author} {\bibfnamefont {X.~Q.}\ \bibnamefont {Huang}}, \bibinfo
  {author} {\bibfnamefont {Y.~M.}\ \bibnamefont {Liu}}, \bibinfo {author}
  {\bibfnamefont {M.}~\bibnamefont {Wang}}, \bibinfo {author} {\bibfnamefont
  {A.}~\bibnamefont {Hu}}, \ and\ \bibinfo {author} {\bibfnamefont {S.~S.}\
  \bibnamefont {Jiang}},\ }\href {http://stacks.iop.org/0295-5075/63/i=6/a=853}
  {\bibfield  {journal} {\bibinfo  {journal} {Europhys. Lett.}\ }\textbf
  {\bibinfo {volume} {63}},\ \bibinfo {pages} {853} (\bibinfo {year}
  {2003})}\BibitemShut {NoStop}%
\bibitem [{\citenamefont {Dal~Negro}\ \emph {et~al.}(2004)\citenamefont
  {Dal~Negro}, \citenamefont {Stolfi}, \citenamefont {Yi}, \citenamefont
  {Michel}, \citenamefont {Duan}, \citenamefont {Kimerling}, \citenamefont
  {LeBlanc},\ and\ \citenamefont {Haavisto}}]{DalNegro2004}%
  \BibitemOpen
  \bibfield  {author} {\bibinfo {author} {\bibfnamefont {L.}~\bibnamefont
  {Dal~Negro}}, \bibinfo {author} {\bibfnamefont {M.}~\bibnamefont {Stolfi}},
  \bibinfo {author} {\bibfnamefont {Y.}~\bibnamefont {Yi}}, \bibinfo {author}
  {\bibfnamefont {J.}~\bibnamefont {Michel}}, \bibinfo {author} {\bibfnamefont
  {X.}~\bibnamefont {Duan}}, \bibinfo {author} {\bibfnamefont {L.~C.}\
  \bibnamefont {Kimerling}}, \bibinfo {author} {\bibfnamefont {J.}~\bibnamefont
  {LeBlanc}}, \ and\ \bibinfo {author} {\bibfnamefont {J.}~\bibnamefont
  {Haavisto}},\ }\href {\doibase 10.1063/1.1764602} {\bibfield  {journal}
  {\bibinfo  {journal} {Appl. Phys. Lett.}\ }\textbf {\bibinfo {volume} {84}},\
  \bibinfo {pages} {5186} (\bibinfo {year} {2004})}\BibitemShut {NoStop}%
\bibitem [{\citenamefont {Grigoriev}\ and\ \citenamefont
  {Biancalana}(2010)}]{Grigoriev2010}%
  \BibitemOpen
  \bibfield  {author} {\bibinfo {author} {\bibfnamefont {V.}~\bibnamefont
  {Grigoriev}}\ and\ \bibinfo {author} {\bibfnamefont {F.}~\bibnamefont
  {Biancalana}},\ }\href {\doibase 10.1016/j.photonics.2010.05.002} {\bibfield
  {journal} {\bibinfo  {journal} {Photon. Nanostruct. Fund. Appl.}\ }\textbf
  {\bibinfo {volume} {8}},\ \bibinfo {pages} {285 } (\bibinfo {year}
  {2010})}\BibitemShut {NoStop}%
\bibitem [{\citenamefont {Hsueh}\ \emph {et~al.}(2011)\citenamefont {Hsueh},
  \citenamefont {Wun}, \citenamefont {Lin},\ and\ \citenamefont
  {Cheng}}]{Hsueh2011}%
  \BibitemOpen
  \bibfield  {author} {\bibinfo {author} {\bibfnamefont {W.~J.}\ \bibnamefont
  {Hsueh}}, \bibinfo {author} {\bibfnamefont {S.~J.}\ \bibnamefont {Wun}},
  \bibinfo {author} {\bibfnamefont {Z.~J.}\ \bibnamefont {Lin}}, \ and\
  \bibinfo {author} {\bibfnamefont {Y.~H.}\ \bibnamefont {Cheng}},\ }\href
  {\doibase 10.1364/JOSAB.28.002584} {\bibfield  {journal} {\bibinfo  {journal}
  {J. Opt. Soc. Am. B}\ }\textbf {\bibinfo {volume} {28}},\ \bibinfo {pages}
  {2584} (\bibinfo {year} {2011})}\BibitemShut {NoStop}%
\bibitem [{\citenamefont {Jiang}\ \emph {et~al.}(2005)\citenamefont {Jiang},
  \citenamefont {Zhang}, \citenamefont {Feng}, \citenamefont {Huang},
  \citenamefont {Yi},\ and\ \citenamefont {Joannopoulos}}]{Jiang2005}%
  \BibitemOpen
  \bibfield  {author} {\bibinfo {author} {\bibfnamefont {X.}~\bibnamefont
  {Jiang}}, \bibinfo {author} {\bibfnamefont {Y.}~\bibnamefont {Zhang}},
  \bibinfo {author} {\bibfnamefont {S.}~\bibnamefont {Feng}}, \bibinfo {author}
  {\bibfnamefont {K.~C.}\ \bibnamefont {Huang}}, \bibinfo {author}
  {\bibfnamefont {Y.}~\bibnamefont {Yi}}, \ and\ \bibinfo {author}
  {\bibfnamefont {J.~D.}\ \bibnamefont {Joannopoulos}},\ }\href {\doibase
  10.1063/1.1928317} {\bibfield  {journal} {\bibinfo  {journal} {Appl. Phys.
  Lett.}\ }\textbf {\bibinfo {volume} {86}},\ \bibinfo {eid} {201110} (\bibinfo
  {year} {2005})}\BibitemShut {NoStop}%
\bibitem [{\citenamefont {Kol\'a\ifmmode~\check{r}\else \v{r}\fi{}}\ and\
  \citenamefont {Nori}(1990)}]{Kolar1990}%
  \BibitemOpen
  \bibfield  {author} {\bibinfo {author} {\bibfnamefont {M.}~\bibnamefont
  {Kol\'a\ifmmode~\check{r}\else \v{r}\fi{}}}\ and\ \bibinfo {author}
  {\bibfnamefont {F.}~\bibnamefont {Nori}},\ }\href {\doibase
  10.1103/PhysRevB.42.1062} {\bibfield  {journal} {\bibinfo  {journal} {Phys.
  Rev. B}\ }\textbf {\bibinfo {volume} {42}},\ \bibinfo {pages} {1062}
  (\bibinfo {year} {1990})}\BibitemShut {NoStop}%
\bibitem [{\citenamefont {Wang}\ \emph {et~al.}(2000)\citenamefont {Wang},
  \citenamefont {Grimm},\ and\ \citenamefont {Schreiber}}]{Wang2000}%
  \BibitemOpen
  \bibfield  {author} {\bibinfo {author} {\bibfnamefont {X.}~\bibnamefont
  {Wang}}, \bibinfo {author} {\bibfnamefont {U.}~\bibnamefont {Grimm}}, \ and\
  \bibinfo {author} {\bibfnamefont {M.}~\bibnamefont {Schreiber}},\ }\href
  {\doibase 10.1103/PhysRevB.62.14020} {\bibfield  {journal} {\bibinfo
  {journal} {Phys. Rev. B}\ }\textbf {\bibinfo {volume} {62}},\ \bibinfo
  {pages} {14020} (\bibinfo {year} {2000})}\BibitemShut {NoStop}%
\bibitem [{\citenamefont {Maslovski}\ \emph {et~al.}(2010)\citenamefont
  {Maslovski}, \citenamefont {Morgado}, \citenamefont {Silveirinha},
  \citenamefont {Kaipa},\ and\ \citenamefont {Yakovlev}}]{Maslovski2010}%
  \BibitemOpen
  \bibfield  {author} {\bibinfo {author} {\bibfnamefont {S.~I.}\ \bibnamefont
  {Maslovski}}, \bibinfo {author} {\bibfnamefont {T.~A.}\ \bibnamefont
  {Morgado}}, \bibinfo {author} {\bibfnamefont {M.~G.}\ \bibnamefont
  {Silveirinha}}, \bibinfo {author} {\bibfnamefont {C.~S.~R.}\ \bibnamefont
  {Kaipa}}, \ and\ \bibinfo {author} {\bibfnamefont {A.~B.}\ \bibnamefont
  {Yakovlev}},\ }\href {http://stacks.iop.org/1367-2630/12/i=11/a=113047}
  {\bibfield  {journal} {\bibinfo  {journal} {New J. Phys.}\ }\textbf {\bibinfo
  {volume} {12}},\ \bibinfo {pages} {113047} (\bibinfo {year}
  {2010})}\BibitemShut {NoStop}%
\bibitem [{\citenamefont {Moharam}\ \emph {et~al.}(1995)\citenamefont
  {Moharam}, \citenamefont {Grann}, \citenamefont {Pommet},\ and\ \citenamefont
  {Gaylord}}]{Moharam1995}%
  \BibitemOpen
  \bibfield  {author} {\bibinfo {author} {\bibfnamefont {M.~G.}\ \bibnamefont
  {Moharam}}, \bibinfo {author} {\bibfnamefont {E.~B.}\ \bibnamefont {Grann}},
  \bibinfo {author} {\bibfnamefont {D.~A.}\ \bibnamefont {Pommet}}, \ and\
  \bibinfo {author} {\bibfnamefont {T.~K.}\ \bibnamefont {Gaylord}},\ }\href
  {\doibase 10.1364/JOSAA.12.001068} {\bibfield  {journal} {\bibinfo  {journal}
  {J. Opt. Soc. Am. A}\ }\textbf {\bibinfo {volume} {12}},\ \bibinfo {pages}
  {1068} (\bibinfo {year} {1995})}\BibitemShut {NoStop}%
\bibitem [{\citenamefont {Maci\'a}(2006)}]{Macia2006}%
  \BibitemOpen
  \bibfield  {author} {\bibinfo {author} {\bibfnamefont {E.}~\bibnamefont
  {Maci\'a}},\ }\href {http://stacks.iop.org/0034-4885/69/i=2/a=R03} {\bibfield
   {journal} {\bibinfo  {journal} {Rep. Progr. Phys.}\ }\textbf {\bibinfo
  {volume} {69}},\ \bibinfo {pages} {397} (\bibinfo {year} {2006})}\BibitemShut
  {NoStop}%
\bibitem [{\citenamefont {Dal~Negro}\ and\ \citenamefont
  {Boriskina}(2012)}]{DalNegro2012}%
  \BibitemOpen
  \bibfield  {author} {\bibinfo {author} {\bibfnamefont {L.}~\bibnamefont
  {Dal~Negro}}\ and\ \bibinfo {author} {\bibfnamefont {S.}~\bibnamefont
  {Boriskina}},\ }\href {\doibase 10.1002/lpor.201000046} {\bibfield  {journal}
  {\bibinfo  {journal} {Laser Photon. Rev.}\ }\textbf {\bibinfo {volume} {6}},\
  \bibinfo {pages} {178} (\bibinfo {year} {2012})}\BibitemShut {NoStop}%
\bibitem [{\citenamefont {Castaldi}\ \emph {et~al.}(2012)\citenamefont
  {Castaldi}, \citenamefont {Galdi}, \citenamefont {Al\`u},\ and\ \citenamefont
  {Engheta}}]{Castaldi2012}%
  \BibitemOpen
  \bibfield  {author} {\bibinfo {author} {\bibfnamefont {G.}~\bibnamefont
  {Castaldi}}, \bibinfo {author} {\bibfnamefont {V.}~\bibnamefont {Galdi}},
  \bibinfo {author} {\bibfnamefont {A.}~\bibnamefont {Al\`u}}, \ and\ \bibinfo
  {author} {\bibfnamefont {N.}~\bibnamefont {Engheta}},\ }\href {\doibase
  10.1103/PhysRevLett.108.063902} {\bibfield  {journal} {\bibinfo  {journal}
  {Phys. Rev. Lett.}\ }\textbf {\bibinfo {volume} {108}},\ \bibinfo {pages}
  {063902} (\bibinfo {year} {2012})}\BibitemShut {NoStop}%
\bibitem [{\citenamefont {Born}\ and\ \citenamefont {Wolf}(1999)}]{Born1999}%
  \BibitemOpen
  \bibfield  {author} {\bibinfo {author} {\bibfnamefont {M.}~\bibnamefont
  {Born}}\ and\ \bibinfo {author} {\bibfnamefont {E.}~\bibnamefont {Wolf}},\
  }\href {http://books.google.com.au/books?id=oV80AAAAIAAJ} {\emph {\bibinfo
  {title} {Principles of Optics, 7th Ed.}}}\ (\bibinfo  {publisher} {Cambridge
  University Press, Cambridge, UK},\ \bibinfo {year} {1999})\BibitemShut
  {NoStop}%
\bibitem [{\citenamefont {Lang}(1987)}]{Lang1987}%
  \BibitemOpen
  \bibfield  {author} {\bibinfo {author} {\bibfnamefont {S.}~\bibnamefont
  {Lang}},\ }\href@noop {} {\emph {\bibinfo {title} {Linear Algebra, 3rd
  Ed.}}}\ (\bibinfo  {publisher} {Springer, Berlin},\ \bibinfo {year}
  {1987})\BibitemShut {NoStop}%
\end{thebibliography}

%

\newpage

\begin{table*}[hbt]
\caption{Geometry and parameters as in Fig. \ref{Figure2}. Coefficients $\alpha_l$, $l=0,2,4,6,8$, pertaining to the effective nonlocal relative permittivity ${\hat \varepsilon}_{\parallel}(k_x)$ in (\ref{eq:epar}), for the $n=1$ and $n=3$ iteration orders.}
\begin{ruledtabular}
\begin{tabular}{cccccc}
$n$  & $\alpha_0$ & $\alpha_2$ & $\alpha_4$ & $\alpha_6$ & $\alpha_8$  \\
\hline
1 & $0.409$ & $-0.034d^2$ & $1.136\cdot 10^{-3}d^4$ & $-2.028\cdot10^{-5}d^6$ & $2.254\cdot10^{-7}d^8$ \\
3 & $1.670$ & $-2.227d^2$ & $1.188d^4$ & $-0.339d^6$ & $0.060d^8$
\end{tabular}
\end{ruledtabular}
\label{Table1}
\end{table*}

\begin{table*}[hbt]
\caption{As in Table \ref{Table1}, but coefficients $\beta_l$, $l=0,2,4,6,8$, pertaining to ${\hat \varepsilon}_{\perp}(k_z)$ in (\ref{eq:eperp}).}
\begin{ruledtabular}
\begin{tabular}{cccccc}
$n$  & $\beta_0$ & $\beta_2$ & $\beta_4$ & $\beta_6$ & $\beta_8$  \\
\hline
1 & $-0.220$ & $-0.017d^2$ & $5.58\cdot10^{-4}d^4$ & $-9.921\cdot10^{-6}d^6$ & $-1.127\cdot10^{-7}d^8$ \\
3 & $0.138$ & $0.134d^2$ & $-8.263\cdot10^{-3}d^4$ & $-0.017d^6$ & $-2.71\cdot10^{-3}d^8$
\end{tabular}
\end{ruledtabular}
\label{Table2}
\end{table*}

%

%

%

%

%

%

%

%

%

%

%

%

%

%

%

%

\end{document}